\documentclass[nofootinbib,reprint,superscriptaddress]{revtex4-2}
\usepackage{amsmath}
\usepackage{amsfonts}
\usepackage{amssymb}
\usepackage{bm}
\usepackage{diagbox}
\usepackage{enumitem}
\usepackage{threeparttable}
\usepackage{lineno}

\usepackage{graphicx}
\usepackage[caption=false]{subfig}

\usepackage[colorlinks]{hyperref}
\hypersetup{citecolor=blue,filecolor=blue,linkcolor=blue,urlcolor=blue}
\usepackage[all]{hypcap}

\usepackage[utf8]{inputenc}
\allowdisplaybreaks

\begin{document}

\title{Effect of non-uniform efficiency on higher-order cumulants in heavy-ion collisions}
\author{Fan Si}\affiliation{State Key Laboratory of Particle Detection and Electronics, University of Science and Technology of China, Hefei 230026, China}
\author{Yifei Zhang}\email{ephy@ustc.edu.cn}\affiliation{State Key Laboratory of Particle Detection and Electronics, University of Science and Technology of China, Hefei 230026, China}
\author{Xiaofeng Luo}\email{xfluo@ccnu.edu.cn}\affiliation{Key Laboratory of Quark \& Lepton Physics (MOE) and Institute of Particle Physics, Central China Normal University, Wuhan 430079, China}

\begin{abstract}

We perform a systematic study on the effect of non-uniform track efficiency correction in higher-order cumulant analysis in heavy-ion collisions. Through analytical derivation, we find that the true values of cumulants can be successfully reproduced by the efficiency correction with an average of the realistic detector efficiency for particles with the same charges within each single phase space. The theoretical conclusions are supported by a toy model simulation by tuning the non-uniformity of the efficiency employed in the track-by-track efficiency correction method. The valid averaged efficiency is found to suppress the statistical uncertainties of the reproduced cumulants dramatically. Thus, usage of the averaged efficiency requires a careful study of phase space dependence. This study is important for carrying out precision measurements of higher-order cumulants in heavy-ion collision experiments at present and in future.

\end{abstract}

\maketitle

\section{Introduction}

In nature, quarks and gluons are confined in nucleons by the strong interaction. In the extremely hot and dense conditions created in relativistic heavy-ion collisions, those partons can be liberated via deconfinement to form Quark-Gluon Plasma (QGP)~\cite{McLerran:1986zb,Bass:1998vz}, which is believed to have existed in the early universe around tens of microseconds after the Big Bang. Experimental evidence shows that the QGP is strongly coupled and behaves like a perfect liquid~\cite{Adams:2005dq,Adcox:2004mh,Muller:2012zq,Akiba:2015jwa,Luo:2020pef,Si:2019esg,Adamczyk:2017xur}. Quantum Chromodynamics (QCD) predicts that at a high temperature ($T$) and small baryon chemical potential ($\mu_B$), it is a smooth crossover from the hadronic phase to the partonic phase~\cite{Aoki:2006we}. At relatively lower $T$ and larger $\mu_B$, QCD-based models predict a first-order phase transition between these two phases~\cite{Karsch:1995sy,deForcrand:2002hgr,Endrodi:2011gv}. Thus, there should be a critical point (CP) connecting the crossover region and first-order phase transition line. One of the most important goals of heavy-ion collisions is to search for the possible QCD critical point in the QCD phase diagram with sensitive observables~\cite{BraunMunzinger:2009zz,Fukushima:2010bq,Gupta:2011wh,Ding:2015ona}.

Higher-order cumulants of conserved charges, such as net-baryon ($B$), net-charge ($Q$) and net-strangeness ($S$), are proposed as sensitive observables for the search of the QCD critical point in heavy-ion collisions~\cite{Stephanov:1999zu,Ejiri:2005wq,Stephanov:2008qz,Asakawa:2009aj,Stephanov:2011pb,Li:2019npr}. In the first phase of the RHIC Beam Energy Scan program (BES-I, 2010-2017), experimentally measured fourth-order net-proton fluctuations ($\kappa\sigma^{2}$) in 0-5\% central Au+Au collisions show a non-monotonic energy dependence with a 3.1$\sigma$ significance~\cite{Luo:2015ewa,Luo:2015doi,STAR:2020tga,STAR:2021iop}. However, due to limited event statistics, the statistical uncertainties are still large below $\sqrt{s_\mathrm{NN}}$ = 20 GeV. To confirm the intriguing observation in BES-I, the STAR experiment has conducted the second phase of the Beam Energy Scan program (BES-II) since 2019, which focuses on the energies below $\sqrt{s_\mathrm{NN}}$ = 20 GeV, with 10-20 times better statistics than those collected in BES-I~\cite{STAR:2010vob}.

In heavy-ion collision experiments, the detector efficiency in the measurement ($\varepsilon_\mathrm{m}$) represents the probability of a produced particle track recorded correctly by the detector systems and applied in later data analysis, which is determined from the detector acceptance, particle detection probability, and track quality selection efficiency. The $\varepsilon_\mathrm{m}$ is always finite and results in loss of incoming track signals recorded by detectors and in a distortion of the multiplicity distributions. Due to the high sensibility of higher-order fluctuations to the event-by-event distribution shapes, the observed values and statistical uncertainties of higher-order cumulants are strongly affected by the finite $\varepsilon_\mathrm{m}$ effect~\cite{Luo:2011tp,Luo:2014rea,Pandav:2018bdx}, therefore, an efficiency correction should be carefully employed in fluctuation analysis. Moreover, the $\varepsilon_\mathrm{m}$ is usually non-uniform and could depend on many factors, such as the collision centrality, track kinematic parameters (transverse momentum, azimuth angle and rapidity) and track crossing effect~\cite{Anderson:2003ur}. For example, the $\varepsilon_\mathrm{m}$ in azimuth can be non-uniform due to an absence of detector acceptance in dead areas or bad electronic readout channels. The non-uniformity of $\varepsilon_\mathrm{m}$ could also contribute to the observed event-by-event fluctuations. However, the efficiency employed in the efficiency correction ($\varepsilon_\mathrm{c}$) is usually an average of the $\varepsilon_\mathrm{m}$ within one or two wide phase spaces (so-called bin-by-bin efficiency)~\cite{Aggarwal:2010wy,Adamczyk:2013dal,Luo:2015ewa}, rather than a realistic non-uniform distribution in various dimensions. In a previous study, a track-by-track-based method has been developed to apply the efficiency correction to the higher-order cumulant analysis of conserved charges~\cite{Luo:2018ofd}. In this paper, we further investigate the effect of non-uniformity in $\varepsilon_\mathrm{c}$, which is important for performing precision measurements of higher-order cumulants in heavy-ion collision experiments at present and in future.

\section{Theoretical study}

In this section, a theoretical study is performed to determine the validity of $\varepsilon_\mathrm{c}$ by tuning its non-uniformity but keeping the same averaged value as $\varepsilon_\mathrm{m}$.

\subsection{Quantity definitions}

In this paper, a phase space is defined to contain $M$ series of particles with the multiplicity vector $\bm{X}=\left(X_1,X_2,\cdots,X_M\right)$, following
\begin{equation}
\label{eq-0}
\tilde{P}(\bm{X})=\sum_{X}P(X)\mathcal{M}_{X,\bm{p}}(\bm{X}),
\end{equation}
where $\bm{p}=\left(p_1,p_2,\cdots,p_M\right)$ and $\mathcal{M}_{X,\bm{p}}(\bm{X})$ denote the multinomial distribution defined as
\begin{equation}
\mathcal{M}_{X,\bm{p}}(\bm{X})=\frac{X!}{\displaystyle{\prod_{i=1}^M X_i!}}\prod_{i=1}^M p_i^{X_i}
\end{equation}
with $\displaystyle{X=\sum_{i=1}^M X_i}$ and $\displaystyle{p_i=\frac{\left\langle X_i\right\rangle}{\langle X\rangle}}$, in other words, particles in a phase space are produced with the total multiplicity $X$ determined by a certain distribution $P(X)$ and allocated into each series by a certain probability vector $\bm{p}$.
For example, if there are independent $X_i\sim\mathrm{Poisson}\left(\lambda_i\right)$, the equation holds for $\displaystyle{X\sim\mathrm{Poisson}\left(\sum\lambda_i\right)}$.

In each event of heavy-ion collisions, there may be several phase spaces in which the particles are usually produced through different physical effects. For example, two phase spaces are considered with two series of particles with the multiplicities $X$ and $Y$ as non-negative integer random variables following the probability distribution function $P(X,Y)$. Note that there may be various types of particles with different charges belonging to a series produced in a phase space. The linear combination of $X$ and $Y$ with the coefficients $a$ and $b$ can be expressed by
\begin{equation}
Q_{(a,b)}=aX+bY,
\end{equation}
The cumulants and factorial cumulants of $Q_{(a,b)}$ are given by~\cite{Nonaka:2017kko,Kitazawa:2017ljq}
\begin{align}
\left\langle Q_{(a,b)}^kQ_{(a',b')}^l\right\rangle_\mathrm{c}&=\left.\partial_{(a,b)}^k\partial_{(a',b')}^lK(\theta,\eta)\right|_{\theta=\eta=0},\\
\left\langle Q_{(a,b)}^kQ_{(a',b')}^l\right\rangle_\mathrm{fc}&=\left.\bar\partial_{(a,b)}^k\bar\partial_{(a',b')}^lK_\mathrm{f}(s,t)\right|_{s=t=1}
\end{align}
from the generating functions
\begin{align}
K(\theta,\eta)&=\ln\sum_{X,Y}P(X,Y)\mathrm{e}^{\theta X+\eta Y},\\
K_\mathrm{f}(s,t)&=\ln\sum_{X,Y}P(X,Y)s^Xt^Y
\end{align}
with
\begin{align}
\partial_{(a,b)}&=a\frac{\partial}{\partial\theta}+b\frac{\partial}{\partial\eta},\\
\bar\partial_{(a,b)}&=a\frac{\partial}{\partial s}+b\frac{\partial}{\partial t}.
\end{align}

The relation between cumulants and factorial cumulants has been derived in Ref.~\cite{Nonaka:2017kko}.

The two phase spaces are divided into $M$ and $N$ bins, respectively, for example, according to the particle types and kinematic parameters. A single particle is allocated into various bins with the probability vector $\bm{p}=\left(p_1,p_2,\cdots,p_M\right)$ or $\bm{q}=\left(q_1,q_2,\cdots,q_N\right)$, which can be determined by the kinematic distributions of particles of different types. Thus, the bin-by-bin particle numbers, represented by the vectors $\bm{X}=\left(X_1,X_2,\cdots,X_M\right)$ and $\bm{Y}=\left(Y_1,Y_2,\cdots,Y_N\right)$, follow the probability distribution function
\begin{equation}
\tilde{P}(\bm{X},\bm{Y})=\sum_{X,Y}P(X,Y)\mathcal{M}_{X,\bm{p}}(\bm{X})\mathcal{M}_{Y,\bm{q}}(\bm{Y}).
\end{equation}

The linear combination of the bin-by-bin particle numbers with the coefficients $\bm{a}=\left(a_1,a_2,\cdots,a_M\right)$ and $\bm{b}=\left(b_1,b_2,\cdots,b_N\right)$ and its cumulants and factorial cumulants are given by
\begin{align}
Q_{(\bm{a},\bm{b})}&=\bm{a}\cdot\bm{X}+\bm{b}\cdot\bm{Y},\\
\left\langle Q_{(\bm{a},\bm{b})}^kQ_{(\bm{a}',\bm{b}')}^l\right\rangle_\mathrm{c}&=\left.\partial_{(\bm{a},\bm{b})}^k\partial_{(\bm{a}',\bm{b}')}^l\tilde{K}(\bm{\theta},\bm{\eta})\right|_{\bm{\theta}=\bm{\eta}=0},\\
\left\langle Q_{(\bm{a},\bm{b})}^kQ_{(\bm{a}',\bm{b}')}^l\right\rangle_\mathrm{fc}&=\left.\bar\partial_{(\bm{a},\bm{b})}^k\bar\partial_{(\bm{a}',\bm{b}')}^l\tilde{K}_\mathrm{f}(\bm{s},\bm{t})\right|_{\bm{s}=\bm{t}=1}
\end{align}
from the generating functions
\begin{align}
\tilde{K}(\bm{\theta},\bm{\eta})&=\ln\sum_{\bm{X},\bm{Y}}\tilde{P}(\bm{X},\bm{Y})\mathrm{e}^{\bm{\theta}\cdot\bm{X}+\bm{\eta}\cdot\bm{Y}},\\
\tilde{K}_\mathrm{f}(\bm{s},\bm{t})&=\ln\sum_{\bm{X},\bm{Y}}\tilde{P}(\bm{X},\bm{Y})\prod_{i=1}^Ms_i^{X_i}\prod_{j=1}^Nt_j^{Y_j}
\end{align}
with
\begin{align}
\partial_{(\bm{a},\bm{b})}&=\sum_{i=1}^Ma_i\frac{\partial}{\partial\theta_i}+\sum_{j=1}^Nb_j\frac{\partial}{\partial\eta_j},\\
\bar\partial_{(\bm{a},\bm{b})}&=\sum_{i=1}^Ma_i\frac{\partial}{\partial s_i}+\sum_{j=1}^Nb_j\frac{\partial}{\partial t_j}.
\end{align}

$Q_{(\bm{a},\bm{b})}$ gives the conserved charge in the collision system if $\bm{a}$ and $\bm{b}$ denote two series of particle charges in various bins, respectively. For two series of particles with the same charges $a_i\equiv a$ ($i$ = 1, 2, $\cdots$, $M$) and $b_j\equiv b$ ($j$ = 1, 2, $\cdots$, $N$), $Q_{(\bm{a},\bm{b})}=Q_{(a,b)}$. The $k$th-order diagonal cumulant $\left\langle Q_{(\bm{a},\bm{b})}^k\right\rangle_\mathrm{c}$ can simply be marked by $C_k$.

$\tilde{K}_\mathrm{f}(\bm{s},\bm{t})$ can be converted into $K_\mathrm{f}(s,t)$ with
\begin{equation}
\begin{split}
\tilde{K}_\mathrm{f}(\bm{s},\bm{t})&=\ln\sum_{X,Y}P(X,Y)\sum_{\bm{X}}\mathcal{M}_{X,\bm{p}}(\bm{X})\prod_{i=1}^Ms_i^{X_i}\\
&\hphantom{\quad\ }\times\sum_{\bm{Y}}\mathcal{M}_{Y,\bm{q}}(\bm{Y})\prod_{j=1}^Nt_j^{Y_j}\\
&=\ln\sum_{X,Y}P(X,Y)\left(\sum_{i=1}^Mp_is_i\right)^X\left(\sum_{j=1}^Nq_jt_j\right)^Y\\
&=K_\mathrm{f}\left(\sum_{i=1}^Mp_is_i,\sum_{j=1}^Nq_jt_j\right),
\end{split}
\end{equation}
where the third line is obtained by multinomial expansion; thus,
\begin{equation}
\bar\partial_{(\bm{a},\bm{b})}^k\bar\partial_{(\bm{a}',\bm{b}')}^l\tilde{K}_\mathrm{f}(\bm{s},\bm{t})=\bar\partial_{(\bm{a}\cdot\bm{p},\bm{b}\cdot\bm{q})}^k\bar\partial_{(\bm{a}'\cdot\bm{p},\bm{b}'\cdot\bm{q})}^lK_\mathrm{f}(s,t).
\end{equation}

Now the finite $\varepsilon_\mathrm{m}$ effect is taken into consideration. A single particle in each bin is detected independently with the probability vector $\bm{\alpha}=\left(\alpha_1,\alpha_2,\cdots,\alpha_M\right)$ or $\bm{\beta}=\left(\beta_1,\beta_2,\cdots,\beta_N\right)$. As a result, the probability distribution function of the measured bin-by-bin particle numbers $\bm{x}=\left(x_1,x_2,\cdots,x_M\right)$ and $\bm{y}=\left(y_1,y_2,\cdots,y_N\right)$ can be expressed by
\begin{equation}
\tilde{\tilde{P}}(\bm{x},\bm{y})=\sum_{\bm{X},\bm{Y}}\tilde{P}(\bm{X},\bm{Y})\prod_{i=1}^M\mathcal{B}_{X_i,\alpha_i}\left(x_i\right)\prod_{j=1}^N\mathcal{B}_{Y_j,\beta_j}\left(y_j\right),
\end{equation}
where $\mathcal{B}_{X_i,\alpha_i}\left(x_i\right)$ and $\mathcal{B}_{Y_j,\beta_j}\left(y_j\right)$ denote the binomial distributions defined by
\begin{equation}
\mathcal{B}_{X_i,\alpha_i}\left(x_i\right)=\frac{X_i!}{x_i!\left(X_i-x_i\right)!}\alpha_i^{x_i}\left(1-\alpha_i\right)^{X_i-x_i}
\end{equation}
as the binomial case of the multinomial distribution. The linear combination of the measured bin-by-bin particle numbers and its cumulants and factorial cumulants are given by
\begin{align}
q_{(\bm{a},\bm{b})}&=\bm{a}\cdot\bm{x}+\bm{b}\cdot\bm{y},\\
\left\langle q_{(\bm{a},\bm{b})}^kq_{(\bm{a}',\bm{b}')}^l\right\rangle_\mathrm{c}&=\left.\partial_{(\bm{a},\bm{b})}^k\partial_{(\bm{a}',\bm{b}')}^l\tilde{\tilde{K}}(\bm{\theta},\bm{\eta})\right|_{\bm{\theta}=\bm{\eta}=0},\\
\left\langle q_{(\bm{a},\bm{b})}^kq_{(\bm{a}',\bm{b}')}^l\right\rangle_\mathrm{fc}&=\left.\bar\partial_{(\bm{a},\bm{b})}^k\bar\partial_{(\bm{a}',\bm{b}')}^l\tilde{\tilde{K}}_\mathrm{f}(\bm{s},\bm{t})\right|_{\bm{s}=\bm{t}=1}
\end{align}
from the generating functions
\begin{align}
\tilde{\tilde{K}}(\bm{\theta},\bm{\eta})&=\ln\sum_{\bm{x},\bm{y}}\tilde{\tilde{P}}(\bm{x},\bm{y})\mathrm{e}^{\bm{\theta}\cdot\bm{x}+\bm{\eta}\cdot\bm{y}},\\
\tilde{\tilde{K}}_\mathrm{f}(\bm{s},\bm{t})&=\ln\sum_{\bm{x},\bm{y}}\tilde{\tilde{P}}(\bm{x},\bm{y})\prod_{i=1}^Ms_i^{x_i}\prod_{j=1}^Nt_j^{y_j}.
\end{align}

The relation between $\tilde{\tilde{K}}_\mathrm{f}(\bm{s},\bm{t})$ and $\tilde{K}_\mathrm{f}(\bm{s},\bm{t})$ can be derived by
\begin{equation}
\begin{split}
\tilde{\tilde{K}}_\mathrm{f}(\bm{s},\bm{t})&=\ln\sum_{\bm{X},\bm{Y}}\tilde{P}(\bm{X},\bm{Y})\sum_{\bm{x}}\prod_{i=1}^M\mathcal{B}_{X_i,\alpha_i}\left(x_i\right)s_i^{x_i}\\
&\hphantom{\quad\ }\times\sum_{\bm{y}}\prod_{j=1}^N\mathcal{B}_{Y_j,\beta_j}\left(y_j\right)t_j^{y_j}\\
&=\ln\sum_{\bm{X},\bm{Y}}\tilde{P}(\bm{X},\bm{Y})\prod_{i=1}^M\left(\alpha_is_i+\left(1-\alpha_i\right)\right)^{X_i}\\
&\hphantom{\quad\ }\times\prod_{j=1}^N\left(\beta_jt_j+\left(1-\beta_j\right)\right)^{Y_j}\\
&=\tilde{K}_\mathrm{f}(\bm{s}',\bm{t}'),
\end{split}
\end{equation}
where $s'_i=\alpha_is_i+\left(1-\alpha_i\right)$ ($i$ = 1, 2, $\cdots$, $M$) and $t'_j=\beta_jt_j+\left(1-\beta_j\right)$ ($j$ = 1, 2, $\cdots$, $N$); thus,
\begin{equation}
\bar\partial_{(\bm{a},\bm{b})}^k\bar\partial_{(\bm{a}',\bm{b}')}^l\tilde{\tilde{K}}_\mathrm{f}(\bm{s},\bm{t})=\bar\partial_{(\bm{a}\bm{\alpha},\bm{b}\bm{\beta})}^k\bar\partial_{(\bm{a}'\bm{\alpha},\bm{b}'\bm{\beta})}^l\tilde{K}_\mathrm{f}(\bm{s},\bm{t}).
\end{equation}
Here, we define the expression
\begin{equation}
\bar\partial_{(\bm{a}\bm{\alpha}/\bm{\alpha}',\bm{b}\bm{\beta}/\bm{\beta}')}=\sum_{i=1}^M\frac{a_i\alpha_i}{\alpha'_i}\frac{\partial}{\partial s_i}+\sum_{j=1}^N\frac{b_j\beta_j}{\beta'_j}\frac{\partial}{\partial t_j},
\end{equation}
which is similar to $Q_{(\bm{a}\bm{\alpha}/\bm{\alpha}',\bm{b}\bm{\beta}/\bm{\beta}')}$, $q_{(\bm{a}\bm{\alpha}/\bm{\alpha}',\bm{b}\bm{\beta}/\bm{\beta}')}$, and $\partial_{(\bm{a}\bm{\alpha}/\bm{\alpha}',\bm{b}\bm{\beta}/\bm{\beta}')}$.

In the efficiency correction using the realistic detector efficiency, the relations between the true $\left\langle Q_{(\bm{a},\bm{b})}^kQ_{(\bm{a}',\bm{b}')}^l\right\rangle_\mathrm{c}$ and the measured $\left\langle q_{(\bm{a},\bm{b})}^kq_{(\bm{a}',\bm{b}')}^l\right\rangle_\mathrm{c}$ are derived by three steps:
\begin{center}
\begin{tabular}{ccc}
$\left\langle Q_{(\bm{a},\bm{b})}^kQ_{(\bm{a}',\bm{b}')}^l\right\rangle_\mathrm{c}$ & $\leftrightarrow$ & $\left\langle Q_{(\bm{a},\bm{b})}^kQ_{(\bm{a}',\bm{b}')}^l\right\rangle_\mathrm{fc}$ \\
 & & $\updownarrow$ \\
$\left\langle q_{(\bm{a},\bm{b})}^kq_{(\bm{a}',\bm{b}')}^l\right\rangle_\mathrm{c}$ & $\leftrightarrow$ & $\left\langle q_{(\bm{a},\bm{b})}^kq_{(\bm{a}',\bm{b}')}^l\right\rangle_\mathrm{fc}$
\end{tabular}
\end{center}
For example, the true diagonal cumulants ($C_k^\mathrm{true}$) up to third-order can be obtained from~\cite{Nonaka:2017kko,Kitazawa:2017ljq}
\begin{widetext}
\begin{align}
\label{eq-1}
C_1^\mathrm{true}=\left\langle Q_{(\bm{a},\bm{b})}\right\rangle_\mathrm{c}&=\left\langle Q_{(\bm{a},\bm{b})}\right\rangle_\mathrm{fc}=\left\langle q_{(\bm{a}/\bm{\alpha},\bm{b}/\bm{\beta})}\right\rangle_\mathrm{fc}=\left\langle q_{(\bm{a}/\bm{\alpha},\bm{b}/\bm{\beta})}\right\rangle_\mathrm{c}=C_1^\mathrm{corr}(\bm{\alpha},\bm{\beta};\bm{\alpha},\bm{\beta}),\\
\label{eq-2}
\begin{split}
C_2^\mathrm{true}=\left\langle Q_{(\bm{a},\bm{b})}^{2}\right\rangle_\mathrm{c}&=\left\langle Q_{(\bm{a},\bm{b})}^{2}\right\rangle_\mathrm{fc}+\left\langle Q_{\left(\bm{a}^{2},\bm{b}^{2}\right)}\right\rangle_\mathrm{fc}=\left\langle q_{(\bm{a}/\bm{\alpha},\bm{b}/\bm{\beta})}^{2}\right\rangle_\mathrm{fc}+\left\langle q_{\left(\bm{a}^{2}/\bm{\alpha},\bm{b}^{2}/\bm{\beta}\right)}\right\rangle_\mathrm{fc}\\
&=\left\langle q_{(\bm{a}/\bm{\alpha},\bm{b}/\bm{\beta})}^{2}\right\rangle_\mathrm{c}-\left\langle q_{\left(\bm{a}^{2}/\bm{\alpha}^{2},\bm{b}^{2}/\bm{\beta}^{2}\right)}\right\rangle_\mathrm{c}+\left\langle q_{\left(\bm{a}^{2}/\bm{\alpha},\bm{b}^{2}/\bm{\beta}\right)}\right\rangle_\mathrm{c}=C_2^\mathrm{corr}(\bm{\alpha},\bm{\beta};\bm{\alpha},\bm{\beta}),
\end{split}\\
\label{eq-3}
\begin{split}
C_3^\mathrm{true}=\left\langle Q_{(\bm{a},\bm{b})}^{3}\right\rangle_\mathrm{c}&=\left\langle Q_{(\bm{a},\bm{b})}^{3}\right\rangle_\mathrm{fc}+3\left\langle Q_{(\bm{a},\bm{b})} Q_{\left(\bm{a}^{2},\bm{b}^{2}\right)}\right\rangle_\mathrm{fc}+\left\langle Q_{\left(\bm{a}^{3},\bm{b}^{3}\right)}\right\rangle_\mathrm{fc}\\
&=\left\langle q_{(\bm{a}/\bm{\alpha},\bm{b}/\bm{\beta})}^{3}\right\rangle_\mathrm{fc}+3\left\langle q_{(\bm{a}/\bm{\alpha},\bm{b}/\bm{\beta})}q_{\left(\bm{a}^{2}/\bm{\alpha},\bm{b}^{2}/\bm{\beta}\right)}\right\rangle_\mathrm{fc}+\left\langle q_{\left(\bm{a}^{3}/\bm{\alpha},\bm{b}^{3}/\bm{\beta}\right)}\right\rangle_\mathrm{fc}\\
&=\left\langle q_{(\bm{a}/\bm{\alpha},\bm{b}/\bm{\beta})}^{3}\right\rangle_\mathrm{c}-3\left\langle q_{(\bm{a}/\bm{\alpha},\bm{b}/\bm{\beta})}q_{\left(\bm{a}^{2}/\bm{\alpha}^{2},\bm{b}^{2}/\bm{\beta}^{2}\right)}\right\rangle_\mathrm{c}+2\left\langle q_{\left(\bm{a}^{3}/\bm{\alpha}^{3},\bm{b}^{3}/\bm{\beta}^{3}\right)}\right\rangle_\mathrm{c}\\
&\hphantom{\quad\ }+3\left\langle q_{(\bm{a}/\bm{\alpha},\bm{b}/\bm{\beta})} q_{\left(\bm{a}^{2}/\bm{\alpha},\bm{b}^{2}/\bm{\beta}\right)}\right\rangle_\mathrm{c}-3\left\langle q_{\left(\bm{a}^{3}/\bm{\alpha}^{2},\bm{b}^{3}/\bm{\beta}^{2}\right)}\right\rangle_\mathrm{c}+\left\langle q_{\left(\bm{a}^{3}/\bm{\alpha},\bm{b}^{3}/\bm{\beta}\right)}\right\rangle_\mathrm{c}\\
&=C_3^\mathrm{corr}(\bm{\alpha},\bm{\beta};\bm{\alpha},\bm{\beta}),
\end{split}
\end{align}
\end{widetext}
where $C_k^\mathrm{corr}(\varepsilon_\mathrm{m};\varepsilon_\mathrm{c})$ denotes $C_k$ corrected with $\varepsilon_\mathrm{c}$ from the cumulants of $q_{(\bm{a},\bm{b})}$ measured with $\varepsilon_\mathrm{m}$.

During an efficiency correction with another set of efficiency $\bm{\alpha}'$ and $\bm{\beta}'$, the detector efficiency $\bm{\alpha}$ and $\bm{\beta}$ in the formulae of the relations between $\left\langle Q_{(\bm{a},\bm{b})}^kQ_{(\bm{a}',\bm{b}')}^l\right\rangle_\mathrm{c}$ and $\left\langle q_{(\bm{a},\bm{b})}^kq_{(\bm{a}',\bm{b}')}^l\right\rangle_\mathrm{c}$ are replaced with $\bm{\alpha}'$ and $\bm{\beta}'$, respectively, such that $C_k^\mathrm{corr}(\bm{\alpha},\bm{\beta};\bm{\alpha}',\bm{\beta}')$.

\subsection{Internally averaged efficiency correction}

The averaged efficiency (AE) is defined as an average of the $\varepsilon_\mathrm{m}$ taken in some or all of the bins divided from the phase spaces, which is uniform in these average bins and inherits the same distribution as the $\varepsilon_\mathrm{m}$ elsewhere. The AE maintains the same averaged value, and its non-uniformity decreases as the average bins increase. According to the relation between phase spaces and average bins, the AE is divided into the internally averaged efficiency (IAE) and the externally averaged efficiency (EAE). The EAE denotes the AE with average bins across multiple phase spaces. Conversely, the IAE is defined as an average taken within each single phase space (not across phase spaces). In this subsection, the validity of the IAE correction is studied.

For example, the efficiency employed in the IAE correction is set to
\begin{align}
\bm{\alpha}'&=\left(\bar{\alpha},\cdots,\bar{\alpha},\alpha_{m+1},\cdots,\alpha_M\right),\\
\bm{\beta}'&=\left(\bar{\beta},\cdots,\bar{\beta},\beta_{n+1},\cdots,\beta_N\right),
\end{align}
with
\begin{align}
\bar{\alpha}&=\frac{\displaystyle{\sum_{i=1}^{m}\left\langle x_i\right\rangle}}{\displaystyle{\sum_{i=1}^{m}\left\langle X_i\right\rangle}}=\frac{\displaystyle{\sum_{i=1}^{m}\alpha_i\left\langle X_i\right\rangle}}{\displaystyle{\sum_{i=1}^{m}\left\langle X_i\right\rangle}}=\frac{\displaystyle{\sum_{i=1}^{m}\alpha_ip_i}}{\displaystyle{\sum_{i=1}^{m}p_i}},\\
\bar{\beta}&=\frac{\displaystyle{\sum_{j=1}^{n}\left\langle y_j\right\rangle}}{\displaystyle{\sum_{j=1}^{n}\left\langle Y_j\right\rangle}}=\frac{\displaystyle{\sum_{j=1}^{n}\beta_j\left\langle Y_j\right\rangle}}{\displaystyle{\sum_{j=1}^{n}\left\langle Y_j\right\rangle}}=\frac{\displaystyle{\sum_{j=1}^{n}\beta_jq_j}}{\displaystyle{\sum_{j=1}^{n}q_j}}
\end{align}
as the average of the detector efficiency $\bm{\alpha}$ or $\bm{\beta}$ in the first $m$ or $n$ bins of a single phase space, respectively. One can find that
\begin{align}
\frac{\alpha_i}{\alpha'_i}&=\left\{\begin{aligned}
&\frac{\alpha_i}{\bar{\alpha}},&&i=1,2,\cdots,m\\
&1,&&i=m+1,m+2,\cdots,M
\end{aligned}\right.\\
\frac{\beta_j}{\beta'_j}&=\left\{\begin{aligned}
&\frac{\beta_j}{\bar{\beta}},&&j=1,2,\cdots,n\\
&1,&&j=n+1,n+2,\cdots,N
\end{aligned}\right.
\end{align}
and
\begin{align}
\sum_{i=1}^{m}{\frac{\alpha_i}{\alpha'_i}p_i}&=\sum_{i=1}^{m}p_i\\
\sum_{j=1}^{n}{\frac{\beta_j}{\beta'_j}q_j}&=\sum_{j=1}^{n}q_j.
\end{align}

If the particle charges are the same in all average bins, that is, $a_1=a_2=\cdots=a_m$ and $b_1=b_2=\cdots=b_n$, it can be derived that
\begin{equation}
\begin{split}
&\hphantom{\quad\ }\bar\partial_{(\bm{a}/\bm{\alpha}',\bm{b}/\bm{\beta}')}\tilde{\tilde{K}}_\mathrm{f}(\bm{s},\bm{t})\\
&=\bar\partial_{(\bm{a}\bm{\alpha}/\bm{\alpha}',\bm{b}\bm{\beta}/\bm{\beta}')}\tilde{K}_\mathrm{f}(\bm{s},\bm{t})\\
&=\left(\sum_{i=1}^M\frac{a_i\alpha_i}{\alpha'_i}p_i\frac{\partial}{\partial s}+\sum_{j=1}^N\frac{b_j\beta_j}{\beta'_j}q_j\frac{\partial}{\partial t}\right)K_\mathrm{f}(s,t)\\
&=\left(\sum_{i=1}^Ma_ip_i\frac{\partial}{\partial s}+\sum_{j=1}^Nb_jq_j\frac{\partial}{\partial t}\right)K_\mathrm{f}(s,t)\\
&=\bar\partial_{(\bm{a},\bm{b})}\tilde{K}_\mathrm{f}(\bm{s},\bm{t})\\
&=\bar\partial_{(\bm{a}/\bm{\alpha},\bm{b}/\bm{\beta})}\tilde{\tilde{K}}_\mathrm{f}(\bm{s},\bm{t})
\end{split}
\end{equation}
and so forth,
\begin{equation}
\begin{split}
\left\langle Q_{(\bm{a},\bm{b})}^kQ_{(\bm{a}',\bm{b}')}^l\right\rangle_\mathrm{fc}&=\left\langle q_{(\bm{a}/\bm{\alpha},\bm{b}/\bm{\beta})}^kq_{(\bm{a}'/\bm{\alpha},\bm{b}'/\bm{\beta})}^l\right\rangle_\mathrm{fc}\\
&=\left\langle q_{(\bm{a}/\bm{\alpha}',\bm{b}/\bm{\beta}')}^kq_{(\bm{a}'/\bm{\alpha}',\bm{b}'/\bm{\beta}')}^l\right\rangle_\mathrm{fc},
\end{split}
\end{equation}
which denotes that the IAE $\bm{\alpha}'$ and $\bm{\beta}'$ can also be employed to reproduce the factorial cumulants and even be converted into cumulants as well as the realistic detector efficiency $\bm{\alpha}$ and $\bm{\beta}$. As a check, $C_k^\mathrm{corr}(\bm{\alpha},\bm{\beta};\bm{\alpha}',\bm{\beta}')=C_k^\mathrm{true}$ if all of the $\bm{\alpha}$ and $\bm{\beta}$ in Eqs.~\eqref{eq-1}-\eqref{eq-3} are replaced with $\bm{\alpha}'$ and $\bm{\beta}'$, respectively. Thus, the IAE correction should be valid unless the average of the detector efficiency is taken for particles with different charges.

\subsection{Externally averaged efficiency correction}

In this subsection, the validity of the EAE correction is further discussed.

Simply, particles with the same charge $a=b=1$ are considered and each phase space is divided into one bin with $M=N=1$. The EAE correction applies the AE with an average bin across two phase spaces as
\begin{equation}
\bar{\varepsilon}=\frac{\langle x\rangle+\langle y\rangle}{\langle X\rangle+\langle Y\rangle}=\frac{\alpha\langle X\rangle+\beta\langle Y\rangle}{\langle X\rangle+\langle Y\rangle}.
\end{equation}

By replacing the realistic detector efficiency in Eq.~\eqref{eq-1} with the EAE, the first-order cumulant is corrected with
\begin{equation}
\begin{split}
C_1^\mathrm{corr}(\alpha,\beta;\bar{\varepsilon},\bar{\varepsilon})&=\left\langle\frac{1}{\bar{\varepsilon}}x+\frac{1}{\bar{\varepsilon}}y\right\rangle_\mathrm{fc}=\left\langle\frac{\alpha}{\bar{\varepsilon}}X+\frac{\beta}{\bar{\varepsilon}}Y\right\rangle_\mathrm{fc}\\
&=\langle X+Y\rangle_\mathrm{c}=C_1^\mathrm{true},
\end{split}
\end{equation}
which is successful for the first-order cumulant. However, to correct the second-order cumulant in the same way, the correction is obtained by
\begin{equation}
\begin{split}
&\hphantom{\quad\ } C_2^\mathrm{corr}(\alpha,\beta;\bar{\varepsilon},\bar{\varepsilon})\\
&=\left\langle\left(\frac{1}{\bar{\varepsilon}}x+\frac{1}{\bar{\varepsilon}}y\right)^2\right\rangle_\mathrm{fc}+\left\langle\frac{1}{\bar{\varepsilon}}x+\frac{1}{\bar{\varepsilon}}y\right\rangle_\mathrm{fc}\\
&=\left\langle\left(\frac{\alpha}{\bar{\varepsilon}}X+\frac{\beta}{\bar{\varepsilon}}Y\right)^2\right\rangle_\mathrm{fc}+\left\langle\frac{\alpha}{\bar{\varepsilon}}X+\frac{\beta}{\bar{\varepsilon}}Y\right\rangle_\mathrm{fc}\\
&=\left\langle\left(\frac{\alpha}{\bar{\varepsilon}}X+\frac{\beta}{\bar{\varepsilon}}Y\right)^2\right\rangle_\mathrm{c}+\left\langle\left(\frac{\alpha}{\bar{\varepsilon}}-\frac{\alpha^2}{\bar{\varepsilon}^2}\right)X+\left(\frac{\beta}{\bar{\varepsilon}}-\frac{\beta^2}{\bar{\varepsilon}^2}\right)Y\right\rangle_\mathrm{c}\\
&\not\equiv\left\langle(X+Y)^2\right\rangle_\mathrm{c}=C_2^\mathrm{true},
\end{split}
\end{equation}
which is valid only in a few specific cases, such as that the detector efficiency $\alpha=\beta=1$ and that $X$ and $Y$ follow the distribution in Eq.~\eqref{eq-0} (a single phase space). The corrections for higher-order cumulants are similar to the second-order case. In general, the EAE cannot reproduce cumulants higher than the first order.

The above two subsections fundamentally provide the validity of the AE correction and its requirements, which can give a reasonable explanation for the numerical results of the AE correction under various conditions reported in Ref.~\cite{Nonaka:2017kko}.

\subsection{Non-averaged efficiency correction}

In this subsection, the validity of the efficiency correction is studied by using the non-averaged efficiency (NAE), which is defined by the efficiency with
\begin{itemize}[topsep=0.5\topsep,itemsep=0ex,parsep=0ex,labelwidth=0.5em,leftmargin=\labelwidth+\labelsep-\itemindent]
\item[-] different non-uniformity from any AE,
\item[-] different non-uniformity from the $\varepsilon_\mathrm{m}$,
\item[-] the same averaged value as the $\varepsilon_\mathrm{m}$.
\end{itemize}

First, only one phase space is considered ($Y\equiv0$), which is divided into two bins with $M=2$. Particles with the same charge $a_1=a_2=1$ are considered. The NAE employed in the correction is marked by $\bm{\alpha}'=\left(\alpha'_1,\alpha'_2\right)$, whose averaged value remains the same as the detector efficiency as
\begin{equation}
\alpha=\frac{\alpha_1\left\langle X_1\right\rangle+\alpha_2\left\langle X_2\right\rangle}{\left\langle X_1\right\rangle+\left\langle X_2\right\rangle}=\frac{\alpha'_1\left\langle X_1\right\rangle+\alpha'_2\left\langle X_2\right\rangle}{\left\langle X_1\right\rangle+\left\langle X_2\right\rangle},
\end{equation}
and the efficiency can be rephrased with
\begin{align}
\alpha_1&=\alpha-h\left\langle X_2\right\rangle,\\
\alpha_2&=\alpha+h\left\langle X_1\right\rangle,\\
\alpha'_1&=\alpha-h'\left\langle X_2\right\rangle,\\
\alpha'_2&=\alpha+h'\left\langle X_1\right\rangle.
\end{align}

The NAE correction for the first-order cumulant is expressed by
\begin{equation}
\begin{split}
&\hphantom{\quad\ } C_1^\mathrm{corr}(\bm{\alpha};\bm{\alpha}')\\
&=\left\langle\frac{x_1}{\alpha'_1}+\frac{x_2}{\alpha'_2}\right\rangle_\mathrm{c}\\
&=\left\langle\frac{\alpha_1}{\alpha'_1}X_1+\frac{\alpha_2}{\alpha'_2}X_2\right\rangle_\mathrm{c}\\
&=\left\langle\frac{\alpha-h\left\langle X_2\right\rangle}{\alpha-h'\left\langle X_2\right\rangle}X_1+\frac{\alpha+h\left\langle X_1\right\rangle}{\alpha+h'\left\langle X_1\right\rangle}X_2\right\rangle_\mathrm{c}\\
&=\left\langle X_1+X_2\right\rangle_\mathrm{c}\left(1+\frac{h'\left(h'-h\right)\left\langle X_1\right\rangle\left\langle X_2\right\rangle}{\left(\alpha+h'\left\langle X_1\right\rangle\right)\left(\alpha-h'\left\langle X_2\right\rangle\right)}\right)\\
&\not\equiv\left\langle X_1+X_2\right\rangle_\mathrm{c}=C_1^\mathrm{true},
\end{split}
\end{equation}
which is valid only if $h'=0$, $h'=h$ or $\left\langle X_1\right\rangle\left\langle X_2\right\rangle=0$. The cases $h'=0$ and $h'=h$ represent the efficiency corrections using the AE and the realistic detector efficiency, respectively, rather than the NAE. If $\left\langle X_1\right\rangle\left\langle X_2\right\rangle=0$, there is only one bin in the phase space, which does not meet the requirement $M=2$. The corrections for higher-order cumulants are also invalid.

Second, two phase spaces are considered with $a=b=1$ and $M=N=1$. The averaged values of the detector efficiency and the NAE $\alpha'$ and $\beta'$ are the same as
\begin{equation}
\bar{\varepsilon}=\frac{\alpha\langle X\rangle+\beta\langle Y\rangle}{\langle X\rangle+\langle Y\rangle}=\frac{\alpha'\langle X\rangle+\beta'\langle Y\rangle}{\langle X\rangle+\langle Y\rangle}.
\end{equation}
The efficiency correction is also invalid in the same way, for example,
\begin{equation}
C_1^\mathrm{corr}(\alpha,\beta;\alpha',\beta')\not\equiv C_1^\mathrm{true}.
\end{equation}

With the above two cases considered, the true values of the cumulants cannot be reproduced by the efficiency correction with the NAE.

\section{Toy model analysis with internally averaged efficiency}

\begin{figure*}[htbp]
\centering
\includegraphics[width=0.8\textwidth]{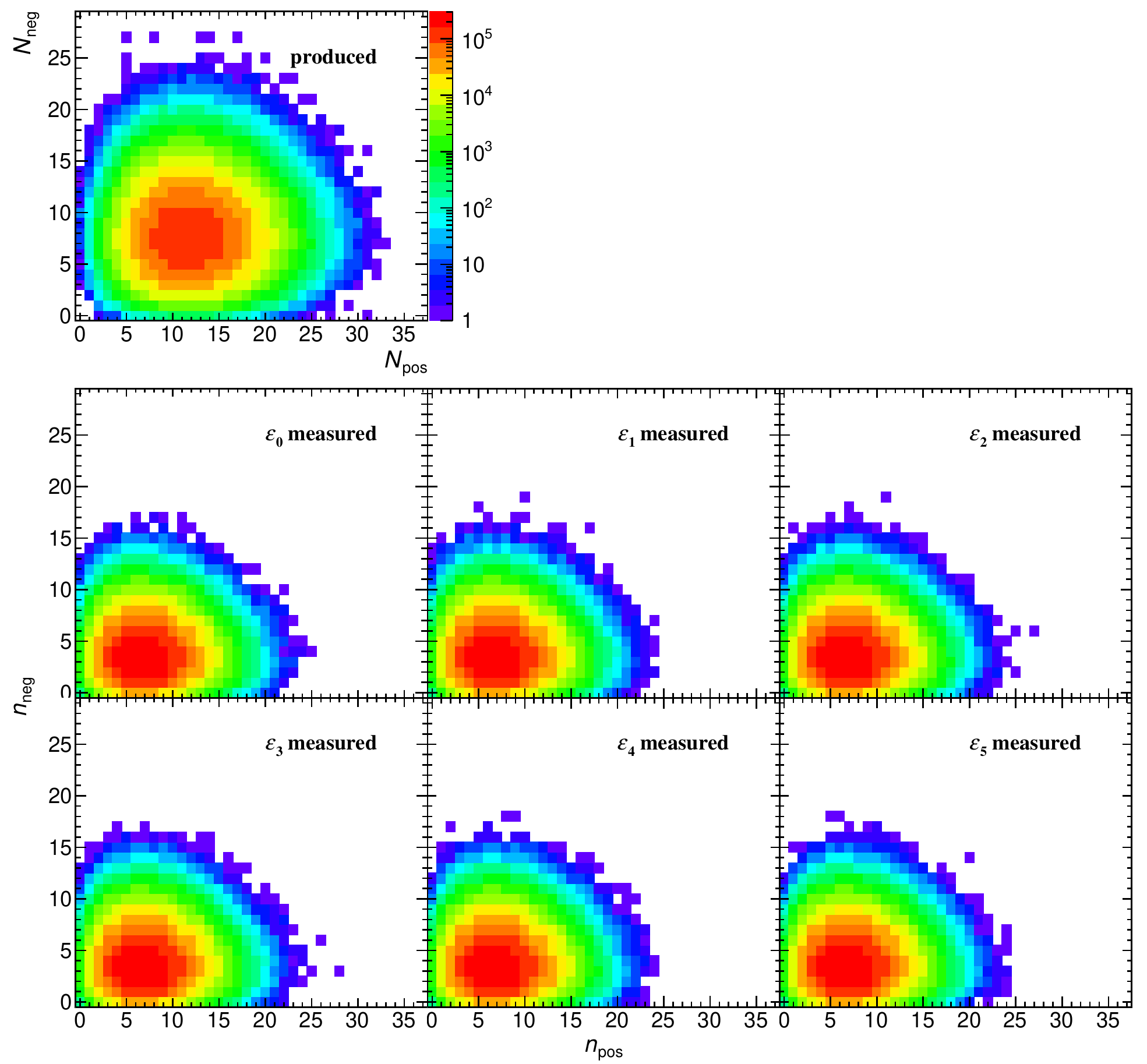}
\caption{(Color online) Correlations of produced and measured numbers of positively and negatively charged particles. The top panel shows the produced distribution, and others show the distributions measured with $\varepsilon_i$ ($i$ = 0, 1, $\cdots$, 5) defined in Eqs.~\eqref{eq12}-\eqref{eq17}, respectively. Due to the equivalence of the averaged values of the efficiency, the measured distributions are indistinguishable.}
\label{fig1}
\end{figure*}

In this section, a toy Monte Carlo model with several sets of IAE with various non-uniformity is employed to check the analytical derivation in the previous section and study the effect of non-uniformity of the $\varepsilon_\mathrm{c}$ on the reproduced cumulants up to fourth-order.

\subsection{Event production}

\begin{figure*}[htbp]
\centering
\includegraphics[width=0.8\textwidth]{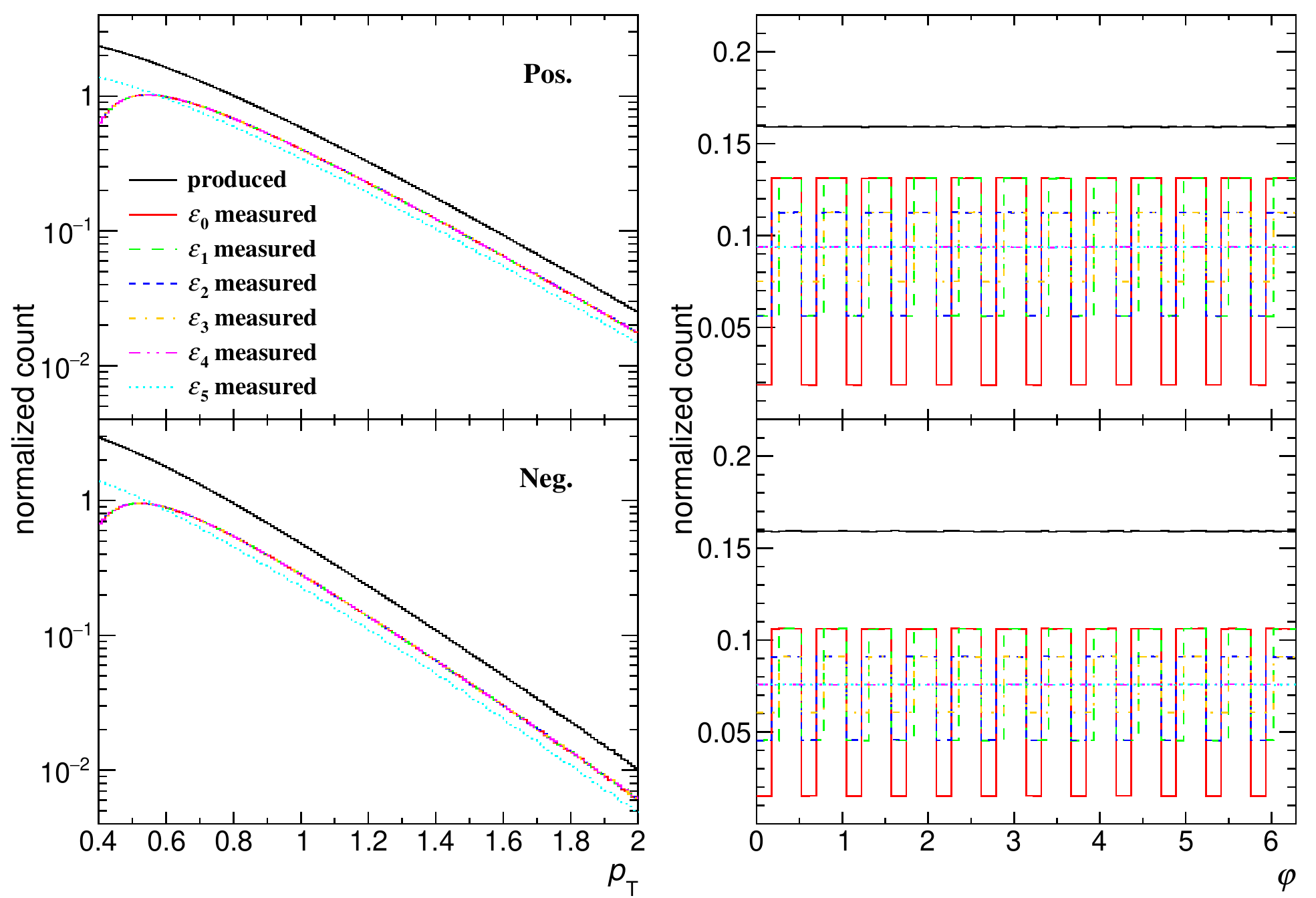}
\caption{(Color online) Distributions of produced and measured $p_{\mathrm T}$ and $\varphi$ of positively (top) and negatively (bottom) charged particles. The produced $p_{\mathrm T}$ and $\varphi$ distributions are shown as black curves, and the colored ones represent the distributions measured with $\varepsilon_i$ ($i$ = 0, 1, $\cdots$, 5) defined in Eqs.~\eqref{eq12}-\eqref{eq17}, respectively. The measured $p_{\mathrm T}$ distributions with $\varepsilon_i$ ($i$ = 0, 1, $\cdots$, 4) are indistinguishable due to their same $u\left(p_\mathrm{T}\right)$ components, and with the same $v(\varphi)$, the measured $\varphi$ distributions with $\varepsilon_4$ and $\varepsilon_5$ almost overlap.}
\label{fig2}
\end{figure*}

\begin{figure*}[htbp]
\centering
\includegraphics[width=0.8\textwidth]{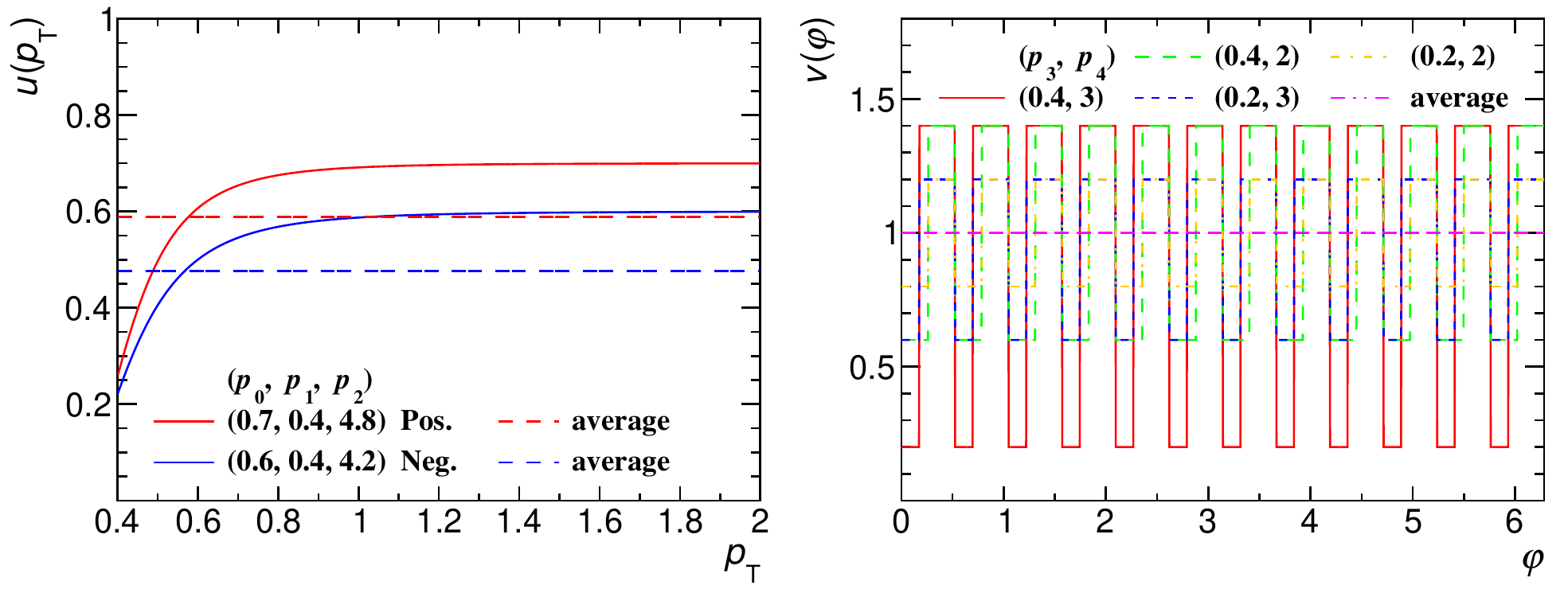}
\caption{(Color online) Two independent one-dimensional functions $u\left(p_\mathrm{T}\right)$ and $v(\varphi)$ defined in Eqs.~\eqref{eq8} and~\eqref{eq9}. The left panel shows $u\left(p_\mathrm{T}\right)$ for positively and negatively charged particles as red and blue solid curves, respectively, and the right panel shows $v(\varphi)$ with various parameters as $\left(p_3,p_4\right)$ = (0.4, 3), (0.4, 2), (0.2, 3), and (0.2, 2). Their averaged values defined in Eqs.~\eqref{eq10} and~\eqref{eq11} are also shown as horizontal dashed lines.}
\label{fig3}
\end{figure*}

We produce $10^7$ events with the numbers of positively and negatively charged particles ($N_\mathrm{pos}$ and $N_\mathrm{neg}$) with the charges $a=1$ and $b=-1$ following two independent Poisson distributions
\begin{align}
\label{eq1}
N_\mathrm{pos}&\sim\mathrm{Poisson}(\lambda_1),\\
N_\mathrm{neg}&\sim\mathrm{Poisson}(\lambda_2),
\label{eq2}
\end{align}
where $\lambda_1$ = 12 and $\lambda_2$ = 8. The 2D distribution of $N_\mathrm{pos}$ and $N_\mathrm{neg}$ is shown in the top panel of Fig.~\ref{fig1}. These particles belong to the same phase space as their multiplicity distributions follow Eq.~\eqref{eq-0}.

Thus, the produced net-charge $N_\mathrm{net}$ (= $N_\mathrm{pos}-N_\mathrm{neg}$) follows the Skellam distribution~\cite{Luo:2014rea}
\begin{equation}
\label{eq3}
N_\mathrm{net}\sim\mathrm{Skellam}(\lambda_1,\lambda_2),
\end{equation}
and the theoretical cumulants up to fourth-order can be calculated as
\begin{align}
\label{eq4}
C_1^{N_\mathrm{net}}=C_3^{N_\mathrm{net}}&=\lambda_1-\lambda_2,\\
C_2^{N_\mathrm{net}}=C_4^{N_\mathrm{net}}&=\lambda_1+\lambda_2.
\label{eq5}
\end{align}

Each of the particles in the produced events is allocated two parameters $p_\mathrm{T}$ and $\varphi$ sampled from
\begin{align}
\label{eq6}
f\left(p_\mathrm{T}\right)&\sim p_\mathrm{T}\mathrm{exp}\left(-p_\mathrm{T}\middle/t\right),0.4\le p_\mathrm{T}<2,\\
g(\varphi)&\sim\mathrm{Uniform}(0, 2\pi),
\label{eq7}
\end{align}
where $t$ = 0.26 and 0.22 for positively and negatively charged particles, respectively. Distributions of the produced $p_\mathrm{T}$ and $\varphi$ are shown as black curves in Fig.~\ref{fig2}. Note that the parameters $p_\mathrm{T}$ and $\varphi$ are abstract parameters in terms of mathematics. Once they are applied to the realistic case in heavy-ion collision experiments, $p_\mathrm{T}$ and $\varphi$ commonly represent the transverse momentum and azimuth angle of the particle with GeV/$c$ and rad as units, respectively. In this case, we ignore the units of $p_\mathrm{T}$ and $\varphi$ in the following discussions.

\subsection{Efficiency definition}

The track efficiency is considered, for example, to be $p_\mathrm{T}$-and-$\varphi$-dependent as a two-dimensional function. Two independent one-dimensional components of efficiency are defined as
\begin{equation}
\label{eq8}
u\left(p_\mathrm{T}\right)=p_0\mathrm{exp}\left(-\left(p_1\middle/p_\mathrm{T}\right)^{p_2}\right),
\end{equation}
where $\left(p_0,p_1,p_2\right)$ = (0.7, 0.4, 4.8) for positively and (0.6, 0.4, 4.2) for negatively
charged particles, respectively, and
\begin{equation}
\label{eq9}
v(\varphi)=\left\{\begin{alignedat}{2}
&1-p_3\left(p_4-1\right),&k\le\left.\varphi\middle/\dfrac{\pi}{6}\right.<k+\dfrac{1}{p_4},\\
&1+p_3,&k+\dfrac{1}{p_4}\le\left.\varphi\middle/\dfrac{\pi}{6}\right.<k+1.
\end{alignedat}\right.(k\in\mathbb{Z})
\end{equation}

Figure~\ref{fig3} shows $u\left(p_\mathrm{T}\right)$ for positively and negatively charged particles as solid curves in the left panel and $v(\varphi)$ with various parameters as $\left(p_3,p_4\right)$ = (0.4, 3), (0.4, 2), (0.2, 3), and (0.2, 2) in the right panel. Their averaged values weighted by the input particle $p_\mathrm{T}$ or $\varphi$ distribution shown as horizontal dashed lines in Fig.~\ref{fig3} are defined as
\begin{align}
\label{eq10}
\mu\left[u\left(p_\mathrm{T}\right)\right]&=\dfrac{\displaystyle{\int_{0.4}^2f\left(p_\mathrm{T}\right)u\left(p_\mathrm{T}\right)\mathrm{d}p_\mathrm{T}}}{\displaystyle{\int_{0.4}^2f\left(p_\mathrm{T}\right)\mathrm{d}p_\mathrm{T}}},\\
\label{eq11}
\mu\left[v(\varphi)\right]&=\dfrac{\displaystyle{\int_0^{2\pi}g(\varphi)v(\varphi)\mathrm{d}\varphi}}{\displaystyle{\int_0^{2\pi}g(\varphi)\mathrm{d}\varphi}}\equiv1,
\end{align}
respectively. There are 12 dips of $v(\varphi)$ with different widths as $p_4$ varies, which are proxies for the low efficiency in 12 $\varphi$ bins due to the dead zone between every two adjacent Time Projection Chamber (TPC) sectors surrounding the heavy-ion beam~\cite{Anderson:2003ur}.

Now, the six sets of track efficiency with different non-uniformity can be expressed as the products of Eqs.~\eqref{eq8} and~\eqref{eq9} given by
\begin{align}
\label{eq12}
\varepsilon_0\left(p_\mathrm{T},\varphi\right)&=u\left(p_\mathrm{T}\right)v(\varphi),\left(p_3,p_4\right)=(0.4,3),\\
\label{eq13}
\varepsilon_1\left(p_\mathrm{T},\varphi\right)&=u\left(p_\mathrm{T}\right)v(\varphi),\left(p_3,p_4\right)=(0.4,2),\\
\label{eq14}
\varepsilon_2\left(p_\mathrm{T},\varphi\right)&=u\left(p_\mathrm{T}\right)v(\varphi),\left(p_3,p_4\right)=(0.2,3),\\
\label{eq15}
\varepsilon_3\left(p_\mathrm{T},\varphi\right)&=u\left(p_\mathrm{T}\right)v(\varphi),\left(p_3,p_4\right)=(0.2,2),\\
\label{eq16}
\varepsilon_4\left(p_\mathrm{T},\varphi\right)&=u\left(p_\mathrm{T}\right),\\
\label{eq17}
\varepsilon_5\left(p_\mathrm{T},\varphi\right)&=\mu\left[u\left(p_\mathrm{T}\right)\right],
\end{align}
whose averaged values and variances are defined as
\begin{align}
\label{eq18}
\begin{split}
\mu\left[\varepsilon_i\left(p_\mathrm{T},\varphi\right)\right]&=\dfrac{\displaystyle{\int_{0.4}^2\int_0^{2\pi}f\left(p_\mathrm{T}\right)g(\varphi)\varepsilon_i\left(p_\mathrm{T},\varphi\right)\mathrm{d}p_\mathrm{T}\mathrm{d}\varphi}}{\displaystyle{\int_{0.4}^2\int_0^{2\pi}f\left(p_\mathrm{T}\right)g(\varphi)\mathrm{d}p_\mathrm{T}\mathrm{d}\varphi}}\\
&=\mu\left[u\left(p_\mathrm{T}\right)\right],
\end{split}\\
\label{eq19}
\sigma^2\left[\varepsilon_i\left(p_\mathrm{T},\varphi\right)\right]&=\mu\left[\varepsilon_i^2\left(p_\mathrm{T},\varphi\right)\right]-\mu^2\left[\varepsilon_i\left(p_\mathrm{T},\varphi\right)\right]
\end{align}
for $i$ = 0, 1, $\cdots$, 5, respectively. The $\mu$ and $\sigma^2$ of the efficiency are not the same for positively and negatively charged particles due to the different $p_\mathrm{T}$ distributions and different parameters of $u\left(p_\mathrm{T}\right)$. Note that the averaged value of $\varepsilon_i\left(p_\mathrm{T},\varphi\right)$, $\mu\left[\varepsilon_i\left(p_\mathrm{T},\varphi\right)\right]$ remains constant for $i$ = 0, 1, $\cdots$, 5. Statistical uncertainties of efficiency-corrected cumulants as a function of AE with a uniform distribution have been studied in Ref.~\cite{Luo:2014rea}. Table~\ref{tab1} summarizes the $\left.\sigma^2\middle/\mu\right.$ values quantifying the non-uniformity of various sets of efficiency. As $i$ ($i$ = 0, 1, $\cdots$, 5) increases, $\left.\sigma^2\middle/\mu\right.$ of $\varepsilon_i$ decreases monotonically to 0, which represents that $\varepsilon_i$ becomes less non-uniform and that $\varepsilon_5$ is completely uniform. 

\begin{table}[htbp]
\centering
\caption{The $\left.\sigma^2\middle/\mu\right.$ values of various sets of efficiency for positively and negatively charged particles.}
\label{tab1}
\begin{tabular}{c|c|c}
\hline
Efficiency & Pos. & Neg. \\
\hline
$\varepsilon_0$ & 0.222 & 0.185 \\
$\varepsilon_1$ & 0.123 & 0.105 \\
$\varepsilon_2$ & 0.074 & 0.065 \\
$\varepsilon_3$ & 0.050 & 0.045 \\
$\varepsilon_4$ & 0.025 & 0.025 \\
$\varepsilon_5$ & 0 & 0 \\
\hline
\end{tabular}
\end{table}

\subsection{Measurement and efficiency correction}
Each particle in the produced events is sampled with six sets of track efficiency, defined in Eqs.~\eqref{eq12}-\eqref{eq17}, as the probability. The 2D distributions of the measured numbers of positively and negatively charged particles ($n_\mathrm{pos}$ and $n_\mathrm{neg}$) are shown in Fig.~\ref{fig1}, and the $p_\mathrm{T}$ and $\varphi$ distributions are shown as colored curves in Fig.~\ref{fig2}.

The so-called track-by-track efficiency corrections~\cite{Luo:2018ofd} with six sets of track efficiency are performed for each of the six measurements to obtain $C_k^\mathrm{corr}\left(\varepsilon_i;\varepsilon_j\right)$ ($i$, $j$ = 0, 1, $\cdots$, 5 and $k$ = 1, 2, 3, 4). Note that $\varepsilon_\mathrm{c}$ may not be the same as $\varepsilon_\mathrm{m}$.

The above procedures, including the production, measurement and efficiency correction, are repeated 1000 times independently. $C_k^\mathrm{corr}$ compared with $C_k^\mathrm{true}$ are shown in Appendix~\ref{appendix}. Here, $\hat{\mu}$ and $\hat{\sigma}$ are defined as the mean value and the standard deviation of $C_k^\mathrm{corr}$, respectively, which can be found at the bottom of each panel of Figs.~\ref{fig12}-\ref{fig17} in Appendix~\ref{appendix}. Figure~\ref{fig4} shows $\chi$ = $\left.\left(\hat{\mu}-C_k^\mathrm{true}\right)\middle/\hat{\sigma}\right.$ for various $\left(\varepsilon_\mathrm{m};\varepsilon_\mathrm{c}\right)$, which quantifies the deviation of $C_k^\mathrm{corr}$ from $C_k^\mathrm{true}$. For $\chi$ far away from $\pm$1, it is drawn as a point at $\pm$1 with a red arrow pointing towards the larger value not shown within the $y$-axis scale, which denotes a failed efficiency correction.

\begin{figure*}[htbp]
\centering
\includegraphics[width=0.8\textwidth]{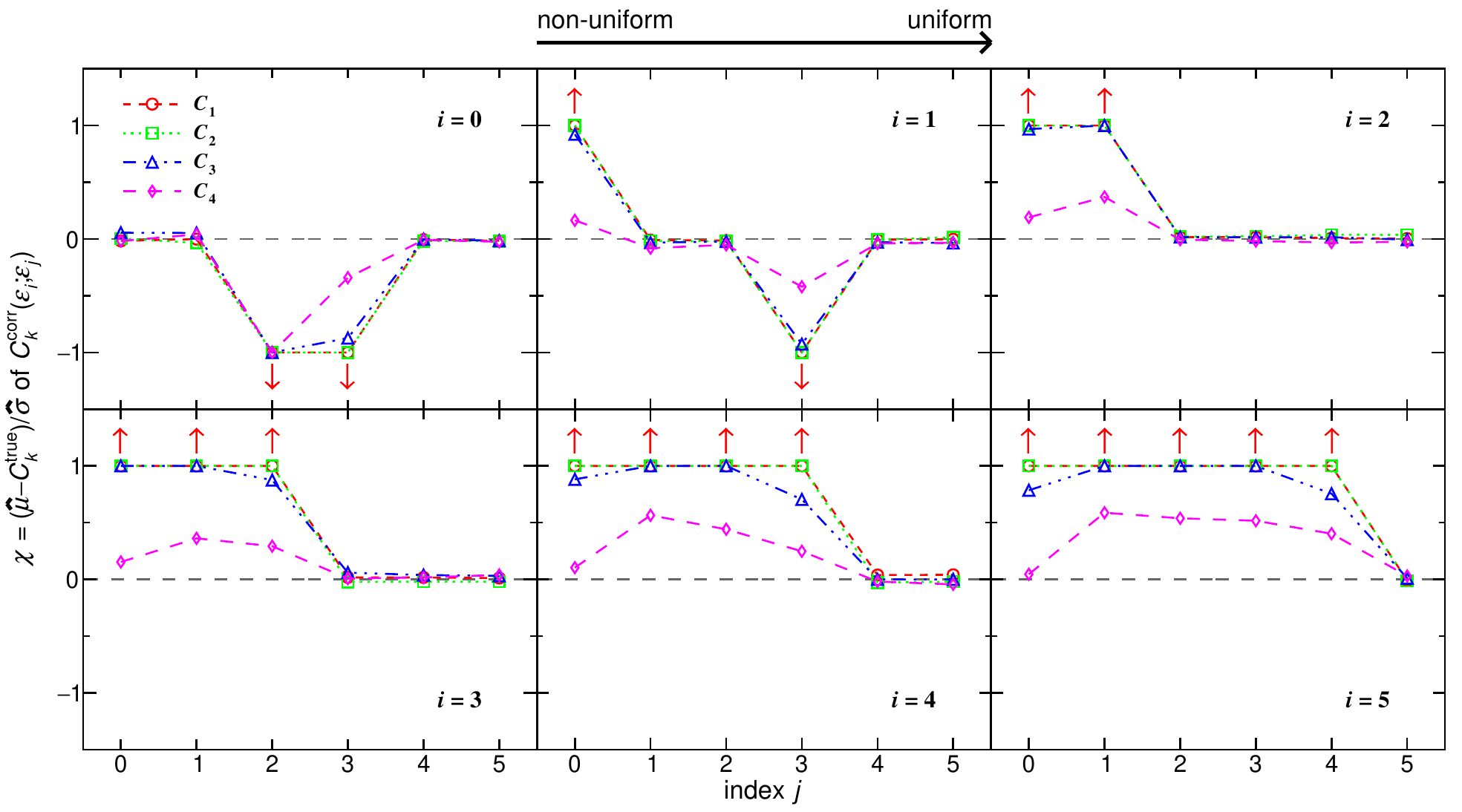}
\caption{(Color online) Quantified deviations $\chi$ = $\left.\left(\hat{\mu}-C_k^\mathrm{true}\right)\middle/\hat{\sigma}\right.$ ($k$ = 1, 2, 3, 4) of the $C_k^\mathrm{corr}$ measured with $\varepsilon_i$ and corrected with $\varepsilon_j$ ($i$, $j$ = 0, 1, $\cdots$, 5) defined in Eqs.~\eqref{eq12}-\eqref{eq17} from $C_k^\mathrm{true}$. For $|\chi|$ $>$ 1 of a failed efficiency correction, the point is drawn at $\pm$1 with a red arrow pointing towards the larger value not shown within the $y$-axis scale.}
\label{fig4}
\end{figure*}

\begin{figure*}[htbp]
\centering
\includegraphics[width=0.8\textwidth]{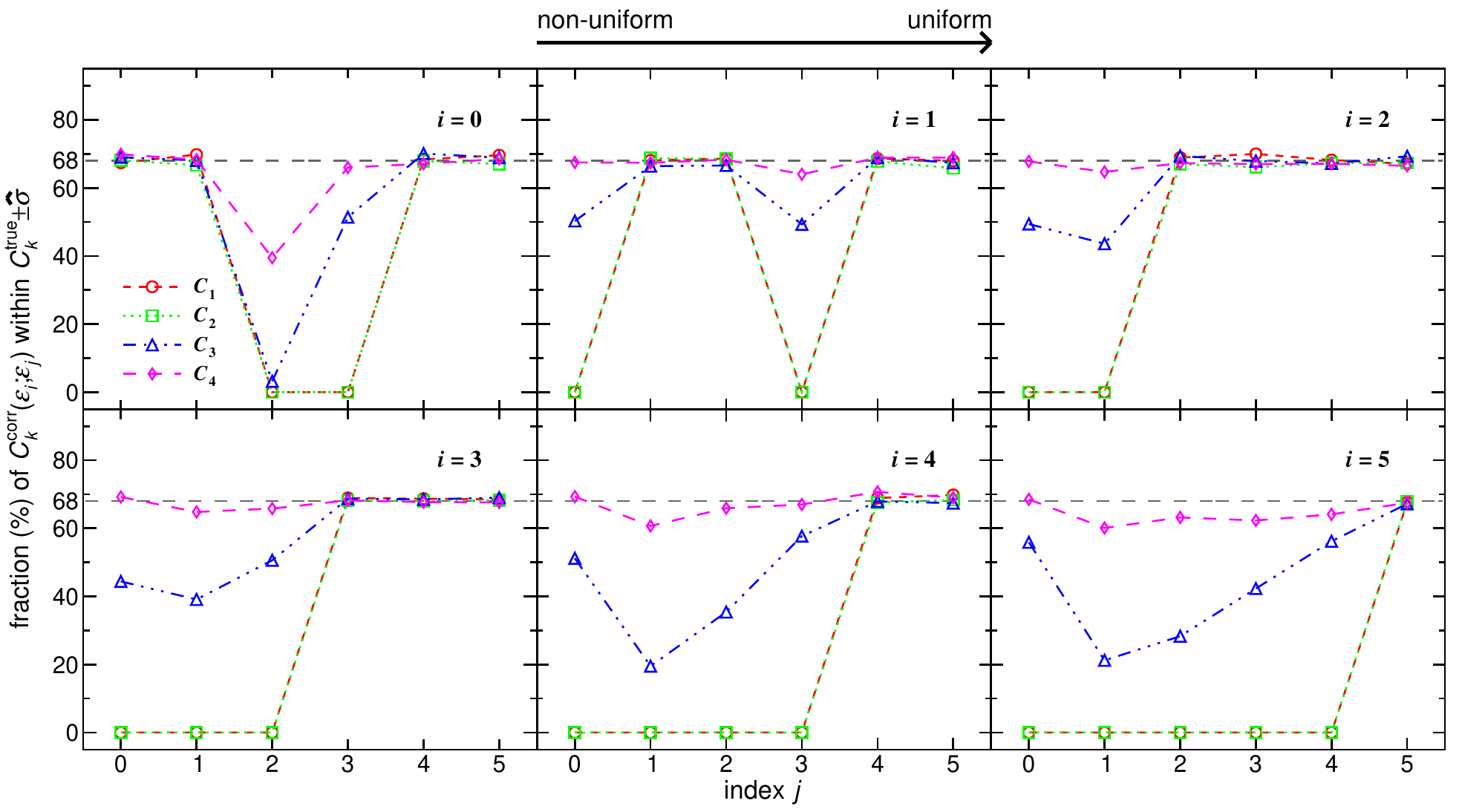}
\caption{(Color online) Fractions of $C_k^\mathrm{corr}$ ($k$ = 1, 2, 3, 4) measured with $\varepsilon_i$ and corrected with $\varepsilon_j$ ($i$, $j$ = 0, 1, $\cdots$, 5) defined in Eqs.~\eqref{eq12}-\eqref{eq17} falling in between $C_k^\mathrm{true}\pm\hat{\sigma}$. The fractions for the cumulants up to fourth-order obtained by each successful efficiency correction should be close to 68\%.}
\label{fig5}
\end{figure*}

Figure~\ref{fig5} shows the fractions of the $C_k^\mathrm{corr}$ falling in between $C_k^\mathrm{true}\pm\hat{\sigma}$ for various $\left(\varepsilon_\mathrm{m};\varepsilon_\mathrm{c}\right)$, which are summarized from the number at the top right of each panel of Figs.~\ref{fig12}-\ref{fig17} in Appendix~\ref{appendix}. All of the fractions of a successful efficiency correction should be close to 68\% as the 1-$\sigma$ probability of the Gaussian distribution.

\begin{figure*}[htbp]
\centering
\includegraphics[width=0.8\textwidth]{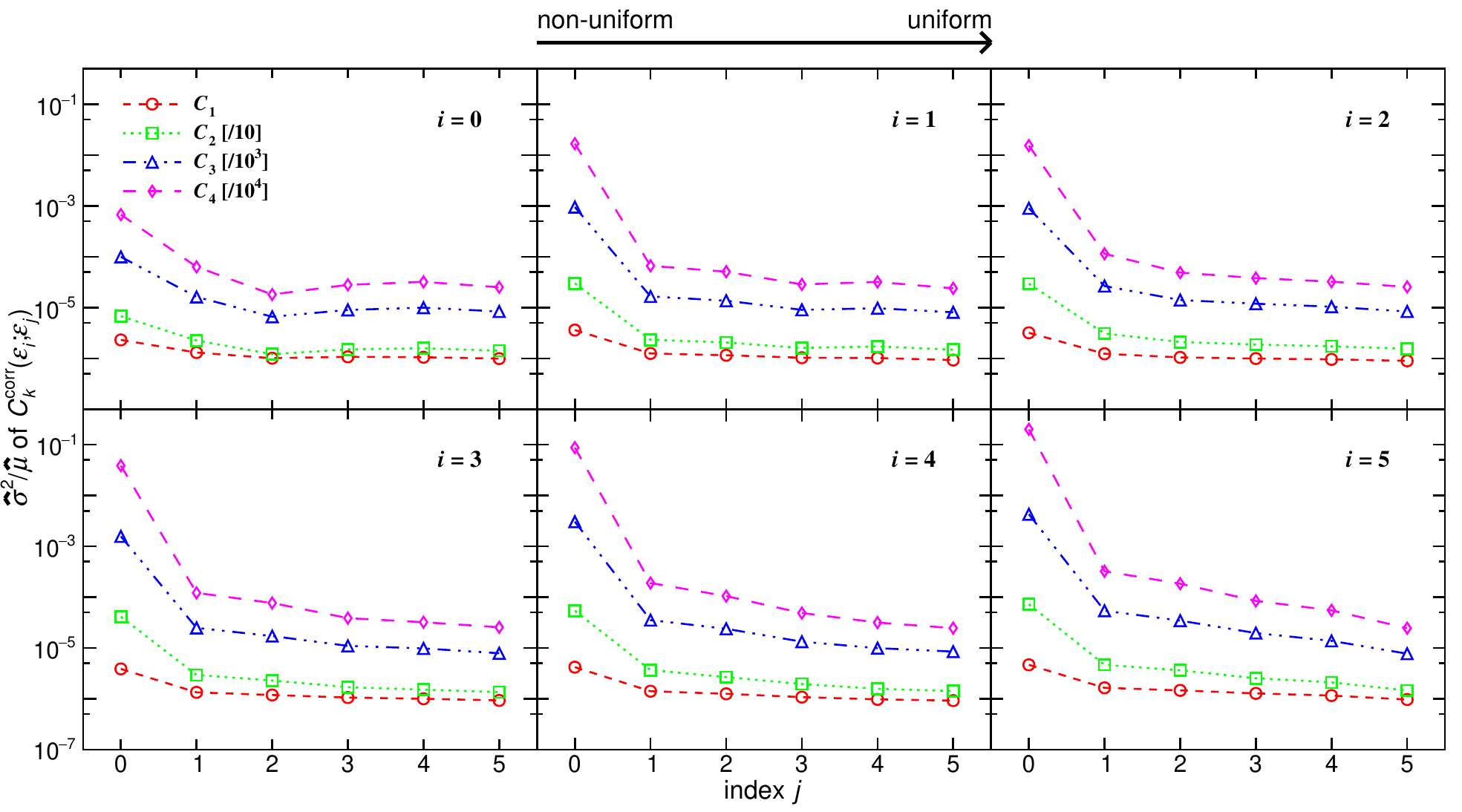}
\caption{(Color online) $\left.\hat{\sigma}^2\middle/\hat{\mu}\right.$ of the $C_k^\mathrm{corr}$ ($k$ = 1, 2, 3, 4) measured with $\varepsilon_i$ and corrected with $\varepsilon_j$ ($i$, $j$ = 0, 1, $\cdots$, 5) defined in Eqs.~\eqref{eq12}-\eqref{eq17}. The values quantify the statistical fluctuations and show a decreasing trend as the non-uniformity of $\varepsilon_\mathrm{c}$ decreases.}
\label{fig6}
\end{figure*}

\begin{table}[htbp]
\centering
\caption{The successful ($\circ$) and failed ($\times$) efficiency corrections for various sets of efficiency employed in measurement ($\varepsilon_\mathrm{m}$) and correction ($\varepsilon_\mathrm{c}$).}
\label{tab2}
\begin{tabular}{c|cccccc}
\hline
\diagbox{$\varepsilon_\mathrm{c}$}{$\varepsilon_\mathrm{m}$} & $\varepsilon_0$ & $\varepsilon_1$ & $\varepsilon_2$ & $\varepsilon_3$ & $\varepsilon_4$ & $\varepsilon_5$ \\
\hline
$\varepsilon_0$ & $\circ$ & $\times$ & $\times$ & $\times$ & $\times$ & $\times$ \\
$\varepsilon_1$ & $\circ$ & $\circ$ & $\times$ & $\times$ & $\times$ & $\times$ \\
$\varepsilon_2$ & $\times$ & $\circ$ & $\circ$ & $\times$ & $\times$ & $\times$ \\
$\varepsilon_3$ & $\times$ & $\times$ & $\circ$ & $\circ$ & $\times$ & $\times$ \\
$\varepsilon_4$ & $\circ$ & $\circ$ & $\circ$ & $\circ$ & $\circ$ & $\times$ \\
$\varepsilon_5$ & $\circ$ & $\circ$ & $\circ$ & $\circ$ & $\circ$ & $\circ$ \\
\hline
\end{tabular}
\end{table}

\begin{figure}[htbp]
\centering
\includegraphics[width=0.8\linewidth]{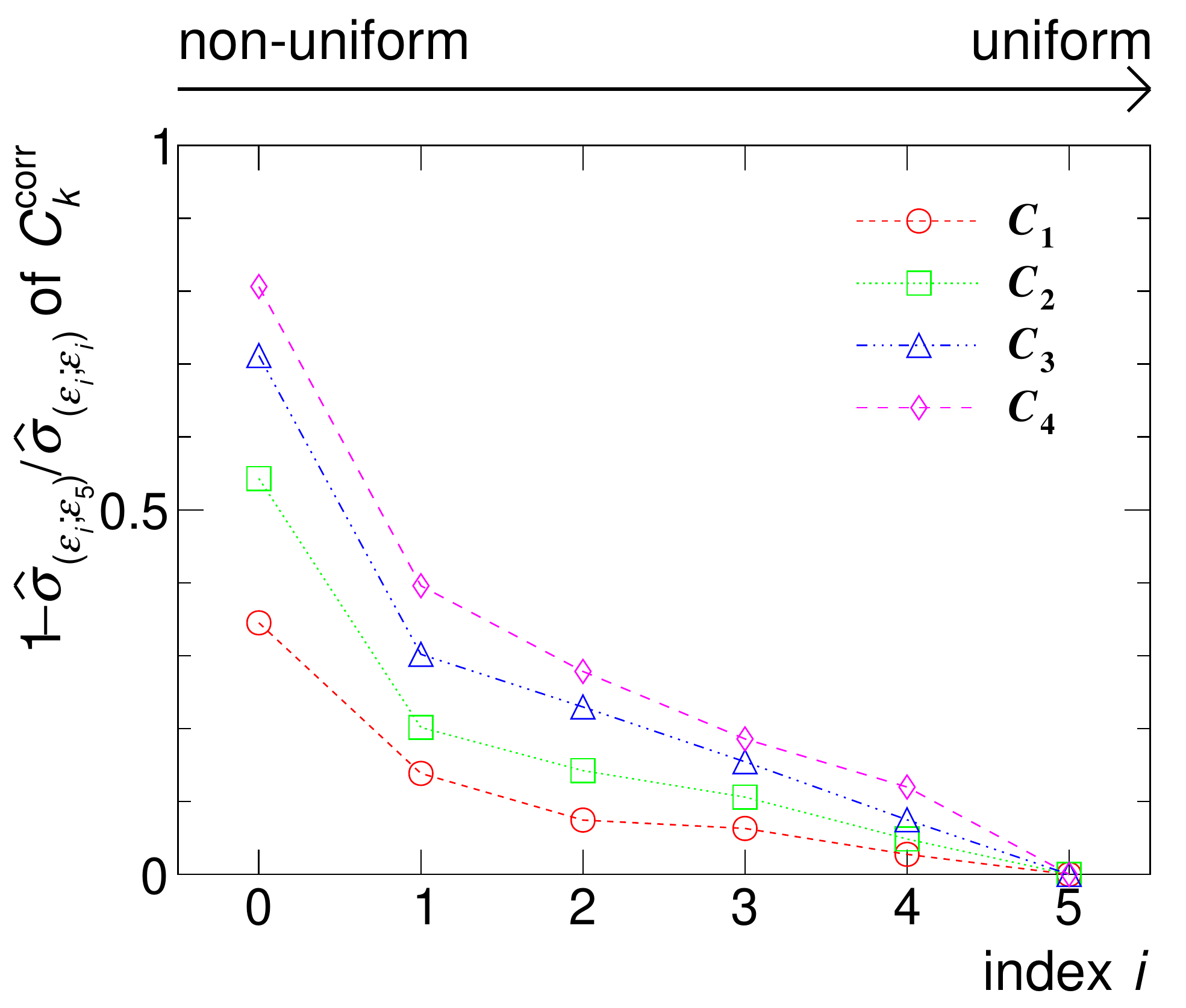}
\caption{(Color online) $1-\left.\hat{\sigma}_{\left(\varepsilon_i;\varepsilon_5\right)}\middle/\hat{\sigma}_{\left(\varepsilon_i;\varepsilon_i\right)}\right.$ of the efficiency-corrected cumulants up to fourth-order for various measurements with $\varepsilon_i$ ($i$ = 0, 1, $\cdots$, 5) defined in Eqs.~\eqref{eq12}-\eqref{eq17}, where $\hat{\sigma}_{\left(\varepsilon_i;\varepsilon_j\right)}$ represents the $\hat{\sigma}$ of the cumulants measured with $\varepsilon_i$ and corrected with $\varepsilon_j$ ($j$ = $i$, 5). The values quantify the suppression of the statistical uncertainties by the uniform efficiency correction and follow a decreasing trend as the non-uniformity of the efficiency employed in the measurement decreases.}
\label{fig7}
\end{figure}

\begin{figure*}[htbp]
\centering
\includegraphics[width=\textwidth]{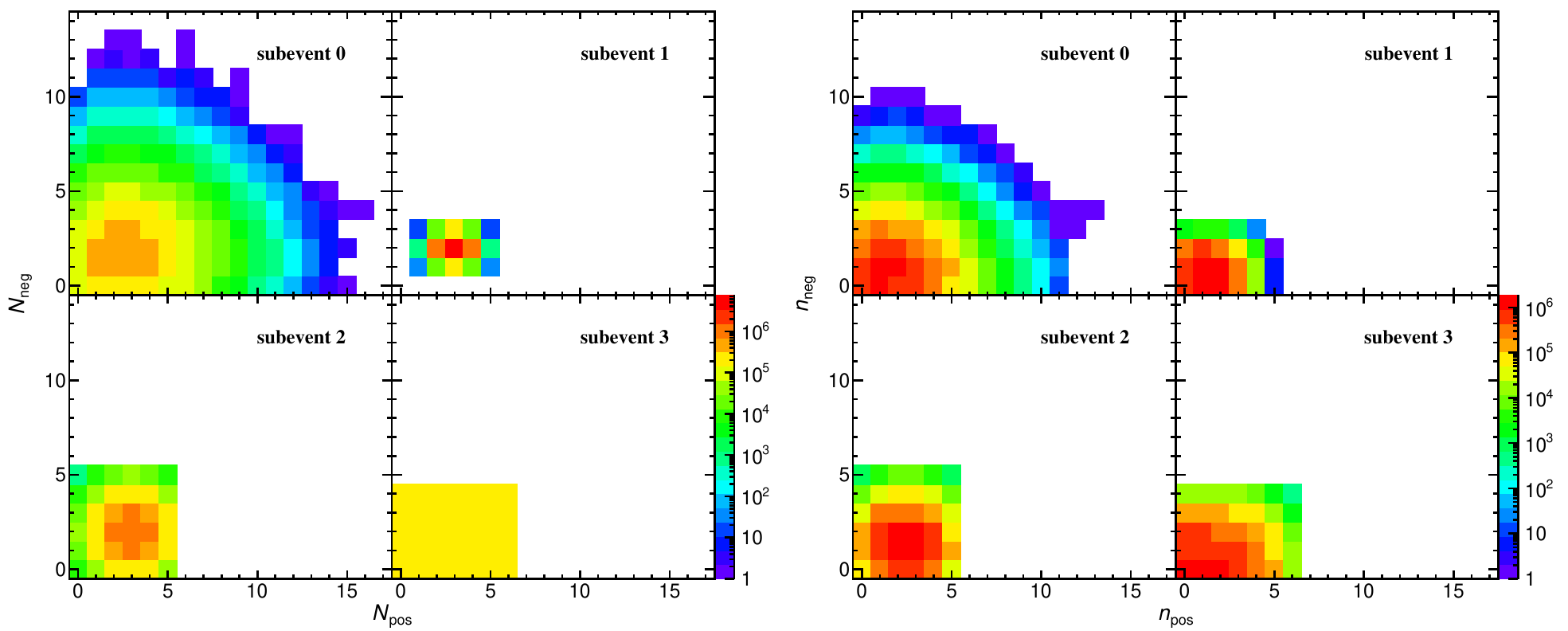}
\caption{(Color online) Correlations of produced and measured numbers of positively and negatively charged particles in various subevents. The left panels show the produced distributions, and the right ones show the distributions measured with $\varepsilon_0$ defined in Eq.~\eqref{eq25}.}
\label{fig8}
\end{figure*}

Table~\ref{tab2} summarizes the successful and failed efficiency corrections for various $\left(\varepsilon_\mathrm{m};\varepsilon_\mathrm{c}\right)$ differentiated by Figs.~\ref{fig4} and~\ref{fig5}, wherein the results show a semi-diagonal. The simulated results support the theoretical predictions presented in the previous section. The diagonal line represents that the $\varepsilon_\mathrm{c}$ is the same as $\varepsilon_\mathrm{m}$, and thus, $C_k^\mathrm{true}$ can be recovered from $C_k^\mathrm{corr}$ within uncertainties, which confirms that the track-by-track efficiency correction method is valid~\cite{Luo:2018ofd}. All of the successful efficiency corrections at the bottom left of Table~\ref{tab2} employ the IAE. For example, $\varepsilon_1$ is the average of $\varepsilon_0$ in two $\varphi$ bins, respectively, as
\begin{equation}
\label{eq20}
\varepsilon_1\left(p_\mathrm{T},\varphi\right)=\left\{\begin{aligned}
&\dfrac{\displaystyle{\int_{k\pi/6}^{(k+1/2)\pi/6}g(\varphi)\varepsilon_0\left(p_\mathrm{T},\varphi\right)\mathrm{d}\varphi}}{\displaystyle{\int_{k\pi/6}^{(k+1/2)\pi/6}g(\varphi)\mathrm{d}\varphi}},\\
&\begin{aligned}k\le\left.\varphi\middle/\dfrac{\pi}{6}\right.<k+\dfrac{1}{2},&\\
\varepsilon_0\left(p_\mathrm{T},\varphi\right),k+\dfrac{1}{2}\le\left.\varphi\middle/\dfrac{\pi}{6}\right.<k+1,&\end{aligned}
\end{aligned}\right.(k\in\mathbb{Z})
\end{equation}
so the correction with $\varepsilon_1$ can be performed successfully for the measurement with $\varepsilon_0$. In the same way, $\varepsilon_4$ is the average of $\varepsilon_{0,1,2,3}$ in the whole $\varphi$ range (0 $\le$ $\varphi$ $<$ 2$\pi$) and thus, $\varepsilon_4$ can be employed to correct the cumulants measured with $\varepsilon_{0,1,2,3}$. However, $\varepsilon_{2,3}$ ($\varepsilon_3$) could be considered as not the IAE of $\varepsilon_0$ ($\varepsilon_1$) but the NAE, which causes three ``$\times$'' at the bottom left of Table~\ref{tab2}. The efficiency corrections by using $\varepsilon_\mathrm{c}$ with more non-uniformity than $\varepsilon_\mathrm{m}$ at the top right of Table~\ref{tab2} also fails for the same reason. Once a non-averaged distribution is induced in the $\varepsilon_\mathrm{c}$ due to incorrect knowledge of the $\varepsilon_\mathrm{m}$, it cannot reproduce the cumulants and results in a failed correction. During the efficiency correction, we should safely take care of the internally averaged track efficiency and avoid introducing additional non-uniformity and non-averaged distributions. However, the statistical uncertainties should be treated carefully as what is studied in the following section.

Figure~\ref{fig6} shows $\left.\hat{\sigma}^2\middle/\hat{\mu}\right.$ for various $\left(\varepsilon_\mathrm{m};\varepsilon_\mathrm{c}\right)$, which quantify the fluctuations of $C_k^\mathrm{corr}$ and strongly relate to their statistical uncertainties. After removing three points of the failed efficiency corrections for $C_k^\mathrm{corr}\left(\varepsilon_0;\varepsilon_2\right)$, $C_k^\mathrm{corr}\left(\varepsilon_0;\varepsilon_3\right)$ and $C_k^\mathrm{corr}\left(\varepsilon_1;\varepsilon_3\right)$, the decreasing trend shows that the statistical fluctuations and uncertainties of the $C_k^\mathrm{corr}$ are dramatically suppressed by the valid IAE corrections with less non-uniformity. In order to quantify how the statistical fluctuations are suppressed by the completely uniform efficiency correction, Fig.~\ref{fig7} summarizes $1-\left.\hat{\sigma}_{\left(\varepsilon_i;\varepsilon_5\right)}\middle/\hat{\sigma}_{\left(\varepsilon_i;\varepsilon_i\right)}\right.$ for various $\varepsilon_\mathrm{m}$, where $\hat{\sigma}_{\left(\varepsilon_i;\varepsilon_5\right)}$ and $\hat{\sigma}_{\left(\varepsilon_i;\varepsilon_i\right)}$ denote the $\hat{\sigma}$ of $C_k^\mathrm{corr}\left(\varepsilon_i;\varepsilon_5\right)$ and $C_k^\mathrm{corr}\left(\varepsilon_i;\varepsilon_i\right)$ ($i$ = 0, 1, $\cdots$, 5), respectively. In the efficiency correction with $\varepsilon_5$, the whole phase spaces are covered by average bins, and the non-uniformity of the track efficiency is ignored, which could flatten the fluctuations and minimize the statistical uncertainties.

In addition, another sample of events are produced with $N_\mathrm{pos}$ and $N_\mathrm{neg}$ following Gaussian distributions as well, and the above procedures are repeated. Similar results are obtained and show the independence of the input particle multiplicity distributions.

\section{Toy model analysis with externally averaged efficiency}

In this section, another toy Monte Carlo model with EAE is further employed in order to check the previous theoretical study about the validity of the EAE correction.

\subsection{Event production}

We produce $10^7$ events with positively and negatively charged particles with the charges $a=1$ and $b=-1$. The events are divided into four subevents according to the particle $p_\mathrm{T}$ and $\varphi$ in various intervals and the multiplicities of two types of particles in four subevents are sampled independently. Table~\ref{tab3} summarizes the subevent divisions and the $N_\mathrm{pos}$ and $N_\mathrm{neg}$ distributions in four subevents. Particles in different subevents belong to different phase spaces. Note that the total numbers of particles are the same in different subevents, as the averaged multiplicities $\left\langle N_\mathrm{pos}\right\rangle\equiv3$ and $\left\langle N_\mathrm{neg}\right\rangle\equiv2$ are constant. The 2D distributions of $N_\mathrm{pos}$ and $N_\mathrm{neg}$ in various subevents are shown in the left panels of Fig.~\ref{fig8}.

\begin{table}[htbp]
\centering
\caption{Subevent divisions and distributions of $N_\mathrm{pos}$ and $N_\mathrm{neg}$ in four subevents.}
\label{tab3}
\begin{threeparttable}
\begin{tabular}{c|cc}
\hline
 & 0 $\le$ $\varphi$ $<$ $\pi$ & $\pi$ $\le$ $\varphi$ $<$ $2\pi$ \\
\hline
0.4 $\le$ $p_\mathrm{T}$ $<$ 0.8 & \begin{tabular}{c}subevent 0\\$N$ $\sim$ Poisson($\lambda$)\end{tabular} & \begin{tabular}{c}subevent 1\\$N$ $\sim$ Gaus($\mu$, $\sigma$)\end{tabular} \\
0.8 $\le$ $p_\mathrm{T}$ $<$ 2 & \begin{tabular}{c}subevent 2\\$N$ $\sim$ Binomial($n$, $p$)\end{tabular} & \begin{tabular}{c}subevent 3\\$N$ $\sim$ Integer($n_\mathrm{max}$)\tnote{*}\end{tabular} \\
\hline
\end{tabular}
\begin{tablenotes}
\footnotesize
\item[*] integer uniformly distributed in the interval [0, $n_\mathrm{max}-1$]
\item $\lambda$ = 3 (pos), 2 (neg)
\item ($\mu$, $\sigma$) = (3, 0.4) (pos), (2, 0.25) (neg)
\item ($n$, $p$) = (5, 0.6) (pos), (5, 0.4) (neg)
\item $n_\mathrm{max}$ = 7 (pos), 5 (neg)
\end{tablenotes}
\end{threeparttable}
\end{table}

\begin{figure*}[htbp]
\centering
\includegraphics[width=0.8\textwidth]{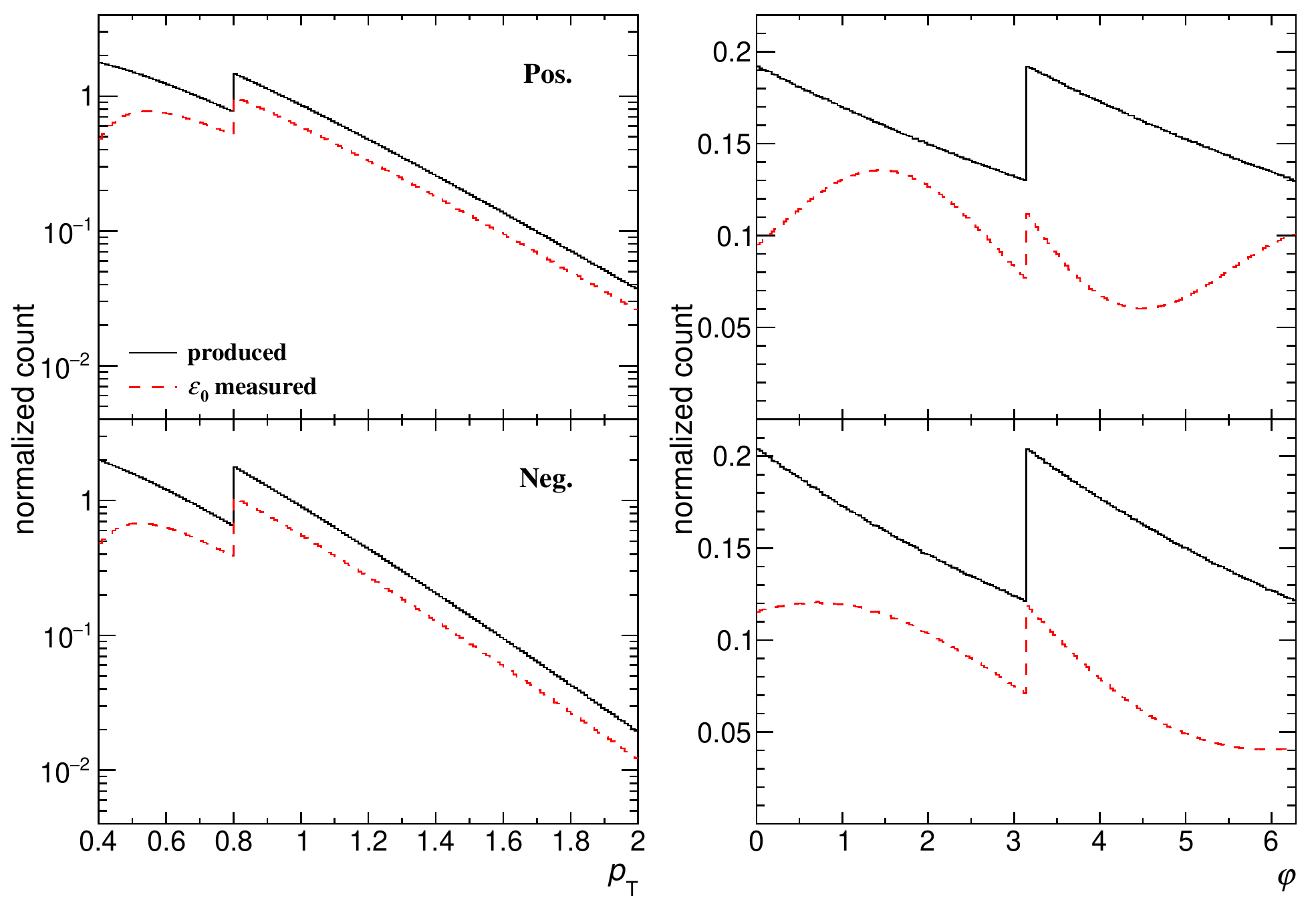}
\caption{(Color online) Distributions of produced and measured $p_{\mathrm T}$ and $\varphi$ of positively (top) and negatively (bottom) charged particles. The produced $p_{\mathrm T}$ and $\varphi$ distributions are shown as black solid curves, and the red dashed ones represent the distributions measured with $\varepsilon_0$ defined in Eq.~\eqref{eq25}. The steps at $p_{\mathrm T}$ = 0.8 and $\varphi$ = $\pi$ are caused by sampling $N_\mathrm{pos}$ and $N_\mathrm{neg}$ from the input distributions individually in multiple phase spaces.}
\label{fig9}
\end{figure*}

\begin{figure*}[htbp]
\centering
\includegraphics[width=0.8\textwidth]{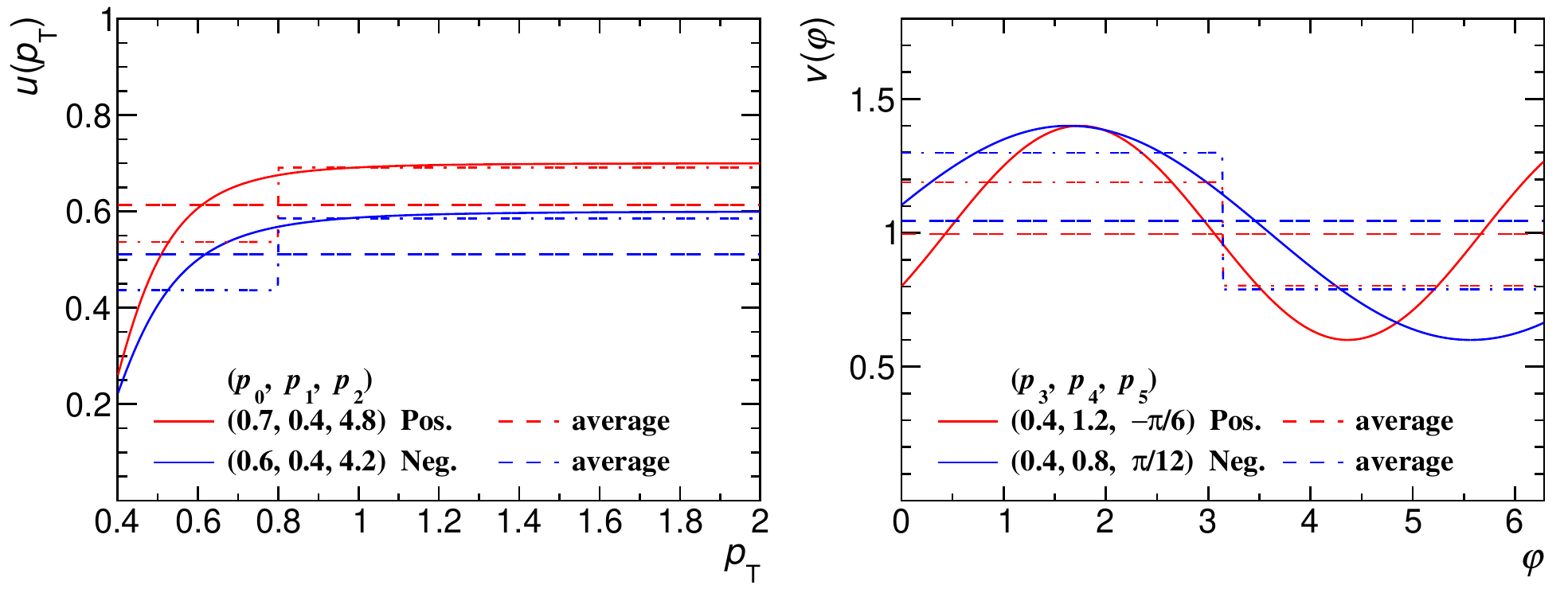}
\caption{(Color online) Two independent one-dimensional functions $u\left(p_\mathrm{T}\right)$ and $v(\varphi)$ defined in Eqs.~\eqref{eq23} and~\eqref{eq24}. The $u\left(p_\mathrm{T}\right)$ and $v(\varphi)$ for positively and negatively charged particles are shown as red and blue solid curves, respectively, and the horizontal dashed lines denote their averaged values in $p_\mathrm{T}$ $\in$ [0.4, 0.8), [0.8, 2) and [0.4, 2) or $\varphi$ $\in$ [0, $\pi$), [$\pi$, $2\pi$) and [0, $2\pi$).}
\label{fig10}
\end{figure*}

Each of the particles in each subevent is allocated $p_\mathrm{T}$ and $\varphi$ sampled from
\begin{align}
\label{eq21}
f\left(p_\mathrm{T}\right)&\sim p_\mathrm{T}\mathrm{exp}\left(-p_\mathrm{T}\middle/t\right),\\
g(\varphi)&\sim\mathrm{exp}(-\varphi/\tau),
\label{eq22}
\end{align}
where $t$ = 0.26 and 0.22 and $\tau$ = 8 and 6 for positively and negatively charged particles, respectively. Distributions of the produced $p_\mathrm{T}$ and $\varphi$ are shown as black solid curves in Fig.~\ref{fig9}.

\subsection{Efficiency definition}

\begin{figure*}[htbp]
\centering
\includegraphics[width=0.8\textwidth]{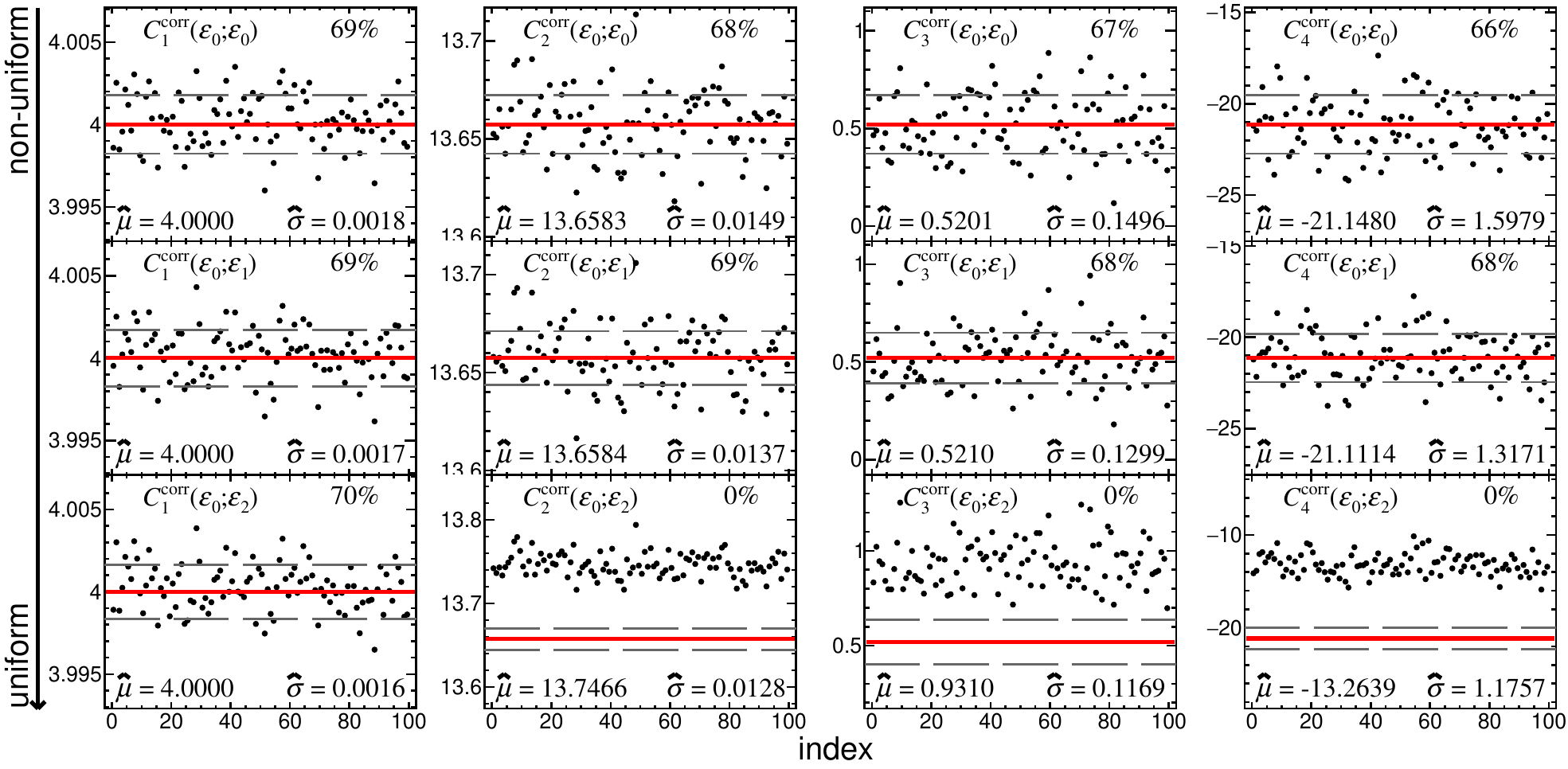}
\caption{(Color online) $C_k^\mathrm{corr}$ ($k$ = 1, 2, 3, 4) measured with $\varepsilon_0$ and corrected with each set of efficiency defined in Eqs.~\eqref{eq25}-\eqref{eq27}. Each panel shows 100 points instead of 1000. $\hat{\mu}$ and $\hat{\sigma}$ at the bottom of each panel show the mean value and the standard deviation of 1000 points. The red solid and gray dashed lines denote $C_k^\mathrm{true}$ and $C_k^\mathrm{true}\pm\hat{\sigma}$, respectively. The number at the top right of each panel represents the fraction of the points falling in between the gray dashed lines.}
\label{fig11}
\end{figure*}

Two independent one-dimensional components of efficiency are defined as
\begin{align}
\label{eq23}
u\left(p_\mathrm{T}\right)&=p_0\mathrm{exp}\left(-\left(p_1\middle/p_\mathrm{T}\right)^{p_2}\right),\\
\label{eq24}
v(\varphi)&=1+p_3\mathrm{sin}\left(p_4\varphi+p_5\right),
\end{align}
where $\left(p_0,p_1,p_2,p_3,p_4,p_5\right)$ = (0.7, 0.4, 4.8, 0.4, 1.2, $-\pi/6$) and (0.6, 0.4, 4.2, 0.4, 0.8, $\pi/12$) for positively and negatively charged particles, respectively. Figure~\ref{fig10} shows these two functions and their averaged values weighted by the input particle $p_\mathrm{T}$ or $\varphi$ distribution in $p_\mathrm{T}$ $\in$ [0.4, 0.8), [0.8, 2) and [0.4, 2) or $\varphi$ $\in$ [0, $\pi$), [$\pi$, $2\pi$) and [0, $2\pi$).

Three sets of track efficiency are given by
\begin{align}
\label{eq25}
\varepsilon_0\left(p_\mathrm{T},\varphi\right)&=u\left(p_\mathrm{T}\right)v(\varphi),\\
\label{eq26}
\varepsilon_1\left(p_\mathrm{T},\varphi\right)&=\mu\left[\varepsilon_0\left(p_\mathrm{T},\varphi\right)\right]_{\mathrm{sub\ of}\left(p_\mathrm{T},\varphi\right)},\\
\label{eq27}
\varepsilon_2\left(p_\mathrm{T},\varphi\right)&=\mu\left[\varepsilon_0\left(p_\mathrm{T},\varphi\right)\right],
\end{align}
where $\mu\left[\varepsilon_0\left(p_\mathrm{T},\varphi\right)\right]_{\mathrm{sub\ of}\left(p_\mathrm{T},\varphi\right)}$ denotes the average of $\varepsilon_0$ within the subevent in which the particle $\left(p_\mathrm{T},\varphi\right)$ falls; therefore, $\varepsilon_1$ is defined as the IAE of $\varepsilon_0$. However, $\varepsilon_2$ represents the EAE as the average of $\varepsilon_0$ across four subevents.

\subsection{Measurement and efficiency correction}

Each particle in the produced events is sampled with $\varepsilon_0$ defined in Eq.~\eqref{eq25} as the probability, and the track-by-track efficiency corrections~\cite{Luo:2018ofd} with the three sets of track efficiency defined in Eqs.~\eqref{eq25}-\eqref{eq27} are performed to obtain $C_k^\mathrm{corr}\left(\varepsilon_0;\varepsilon_j\right)$ ($j$ = 0, 1, 2 and $k$ = 1, 2, 3, 4). The 2D distributions of $n_\mathrm{pos}$ and $n_\mathrm{neg}$ among the four subevents are shown in the right panels of Fig.~\ref{fig8}, and the $p_\mathrm{T}$ and $\varphi$ distributions are shown as red dashed curves in Fig.~\ref{fig9}.

The production, measurement and efficiency correction are independently repeated 1000 times. A comparison of $C_k^\mathrm{corr}$ with $C_k^\mathrm{true}$ is shown in Fig.~\ref{fig11}. $\hat{\mu}$ and $\hat{\sigma}$ of 1000 points are shown at the bottom of each panel. The red solid lines represent $C_k^\mathrm{true}$, and the gray dashed lines denote $C_k^\mathrm{true}\pm\hat{\sigma}$. The fraction of $C_k^\mathrm{corr}$ falling in between the gray dashed lines is shown at the top right of each panel, which should be comparable with 68\% for a valid efficiency correction. Figure~\ref{fig11} clearly shows that $C_k^\mathrm{true}$ can be successfully reproduced by the efficiency correction with $\varepsilon_1$ for the measurement with $\varepsilon_0$ within uncertainties, which supports the validity of the IAE correction. However, there is no $C_2$, $C_3$, or $C_4$ corrected with $\varepsilon_2$ between the gray dashed lines in Fig.~\ref{fig11}, in other words, the EAE correction is not valid for cumulants higher than the first order, which has been predicted by a previous theoretical study. This indicates that the usage of the AE correction requires comprehensive knowledge about the phase space dependence and the EAE correction for higher-order cumulants should be avoided in data analysis.

\section{Summary}

In this paper, we investigate the effect of non-uniform track efficiency on the efficiency correction of higher-order cumulants of conserved charges. This theoretical study proves that an average of the detector efficiency taken for particles with the same charge within each single phase space can successfully reproduce the true values of cumulants by tuning the track efficiency employed in the efficiency correction. The toy model simulation with the track-by-track efficiency correction supports the analytical proof and shows that the valid averaged efficiency correction results in a significant suppression of the statistical uncertainties of cumulants as the event-by-event fluctuations resulting from the non-uniformity of the detector efficiency are not taken into account. However, the correction with the averaged efficiency across multiple phase spaces or the efficiency with a non-uniformity different from that in the averaged case is invalid; a comprehensive knowledge of different phase spaces is required when applying an averaged efficiency correction. If we safely use the realistic detector efficiency in the correction, precisely understanding the cumulants of conserved charges with small uncertainties requires detector efficiency that is as uniform as possible.

The uniformity of the detector design and construction is crucial for precision measurements of higher-order fluctuations. Future fluctuation analysis in the lower energy and higher baryon density regions calls for comprehensive understanding of the uniformity of detectors, such as NICA-MPD~\cite{Golovatyuk:2016zps} and FAIR-CBM~\cite{CBM:2016kpk}.

\acknowledgments{This work was partially supported by National Key Research and Development Program of China with Grant No. 2018YFE0205200, 2018YFE0104700, 2020YFE0202002, Natural Science Foundation of China with Grant No. 11890711, 11890712, 12122505, 11828501, 11861131009, Anhui Provincial Natural Science Foundation with Grant No. 1808085J02 and Innovation Fund of Key Laboratory of Quark and Lepton Physics with Grant No. QLPL2020P01.}

\bibliographystyle{apsrev4-2}
\bibliography{references.bib}

\appendix

\section{}
\label{appendix}

\begin{figure*}[htbp]
\centering
\includegraphics[width=0.8\textwidth]{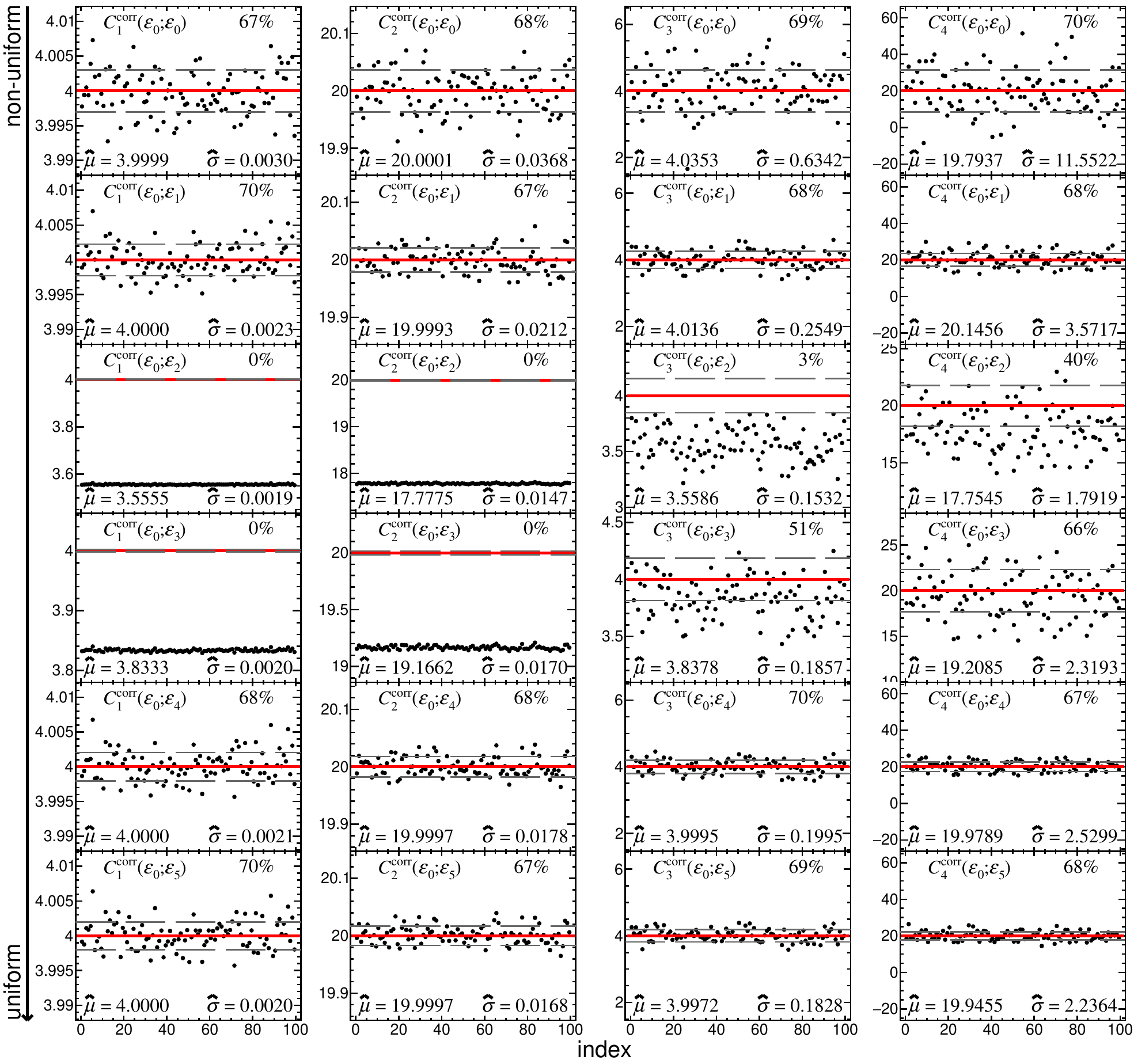}
\caption{(Color online) $C_k^\mathrm{corr}$ ($k$ = 1, 2, 3, 4) measured with $\varepsilon_0$ and corrected with each set of efficiency defined in Eqs.~\eqref{eq12}-\eqref{eq17}. Each panel shows 100 points instead of 1000. $\hat{\mu}$ and $\hat{\sigma}$ at the bottom of each panel show the mean value and the standard deviation of 1000 points. The red solid and gray dashed lines denote $C_k^\mathrm{true}$ and $C_k^\mathrm{true}\pm\hat{\sigma}$, respectively. The number at the top right of each panel represents the fraction of the points falling in between the gray dashed lines.}
\label{fig12}
\end{figure*}

\begin{figure*}[htbp]
\centering
\includegraphics[width=0.8\textwidth]{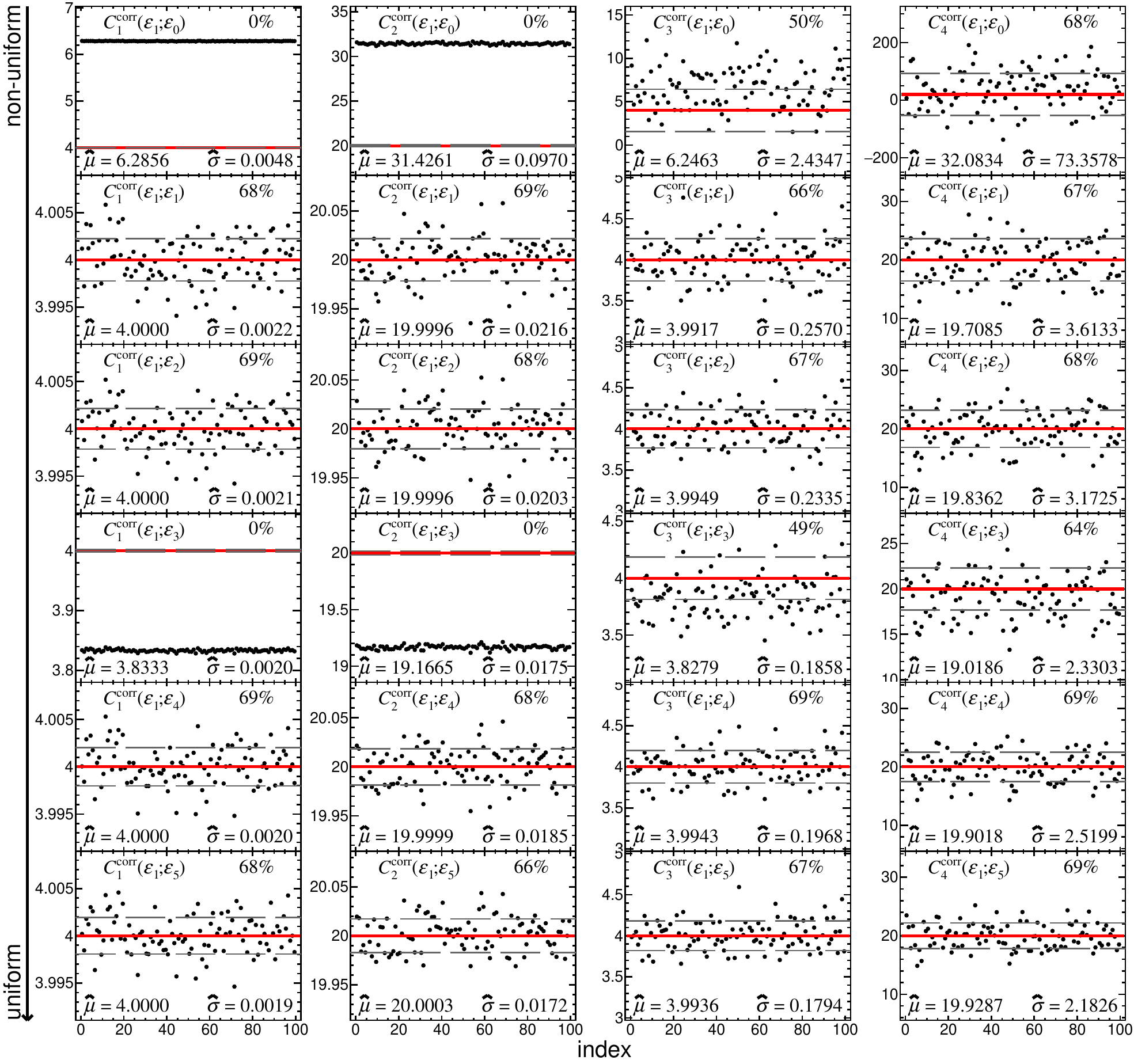}
\caption{(Color online) $C_k^\mathrm{corr}$ ($k$ = 1, 2, 3, 4) measured with $\varepsilon_1$ and corrected with each set of efficiency defined in Eqs.~\eqref{eq12}-\eqref{eq17}. Each panel shows 100 points instead of 1000. $\hat{\mu}$ and $\hat{\sigma}$ at the bottom of each panel show the mean value and the standard deviation of 1000 points. The red solid and gray dashed lines denote $C_k^\mathrm{true}$ and $C_k^\mathrm{true}\pm\hat{\sigma}$, respectively. The number at the top right of each panel represents the fraction of the points falling in between the gray dashed lines.}
\label{fig13}
\end{figure*}

\begin{figure*}[htbp]
\centering
\includegraphics[width=0.8\textwidth]{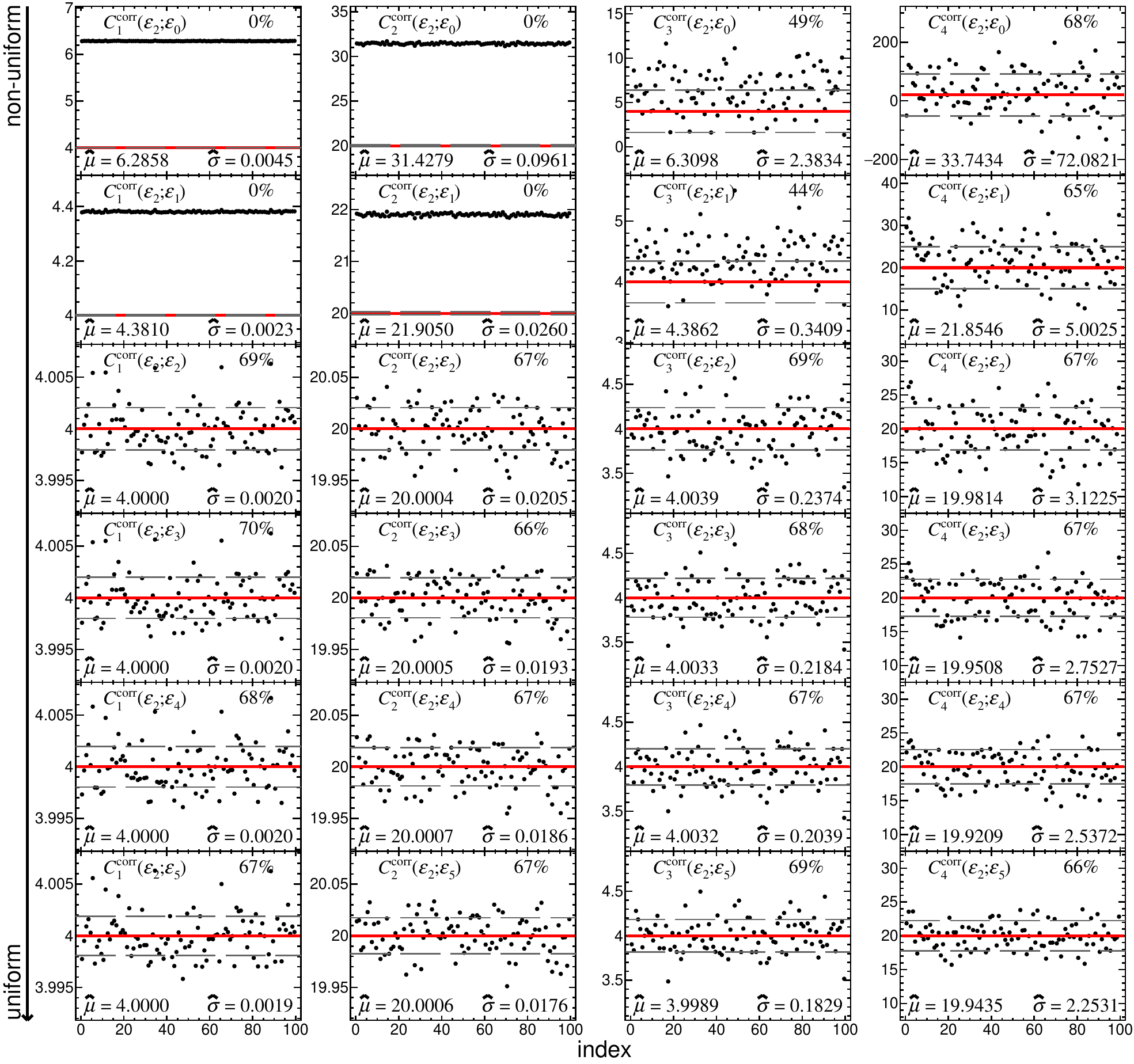}
\caption{(Color online) $C_k^\mathrm{corr}$ ($k$ = 1, 2, 3, 4) measured with $\varepsilon_2$ and corrected with each set of efficiency defined in Eqs.~\eqref{eq12}-\eqref{eq17}. Each panel shows 100 points instead of 1000. $\hat{\mu}$ and $\hat{\sigma}$ at the bottom of each panel show the mean value and the standard deviation of 1000 points. The red solid and gray dashed lines denote $C_k^\mathrm{true}$ and $C_k^\mathrm{true}\pm\hat{\sigma}$, respectively. The number at the top right of each panel represents the fraction of the points falling in between the gray dashed lines.}
\label{fig14}
\end{figure*}

\begin{figure*}[htbp]
\centering
\includegraphics[width=0.8\textwidth]{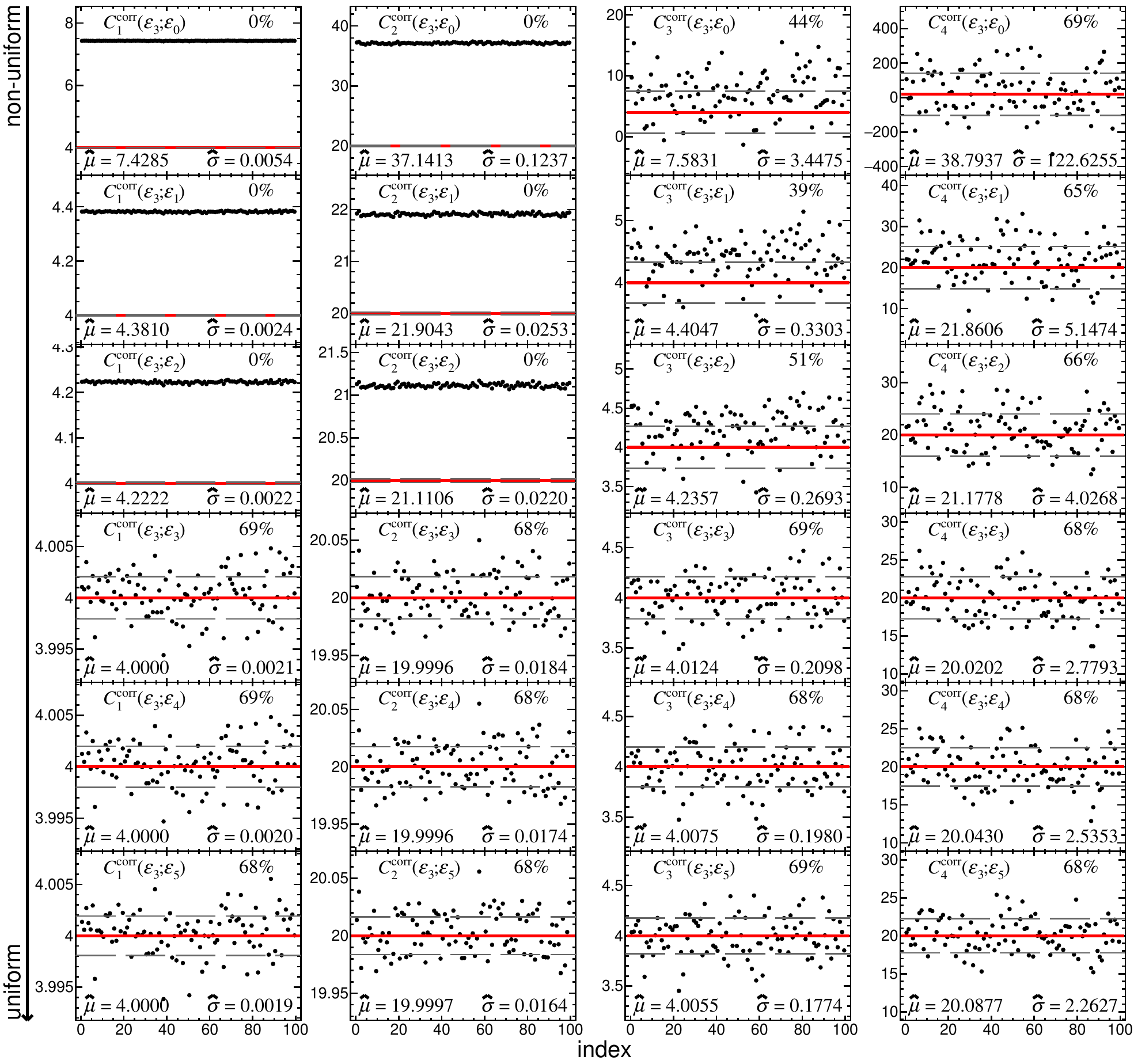}
\caption{(Color online) $C_k^\mathrm{corr}$ ($k$ = 1, 2, 3, 4) measured with $\varepsilon_3$ and corrected with each set of efficiency defined in Eqs.~\eqref{eq12}-\eqref{eq17}. Each panel shows 100 points instead of 1000. $\hat{\mu}$ and $\hat{\sigma}$ at the bottom of each panel show the mean value and the standard deviation of 1000 points. The red solid and gray dashed lines denote $C_k^\mathrm{true}$ and $C_k^\mathrm{true}\pm\hat{\sigma}$, respectively. The number at the top right of each panel represents the fraction of the points falling in between the gray dashed lines.}
\label{fig15}
\end{figure*}

\begin{figure*}[htbp]
\centering
\includegraphics[width=0.8\textwidth]{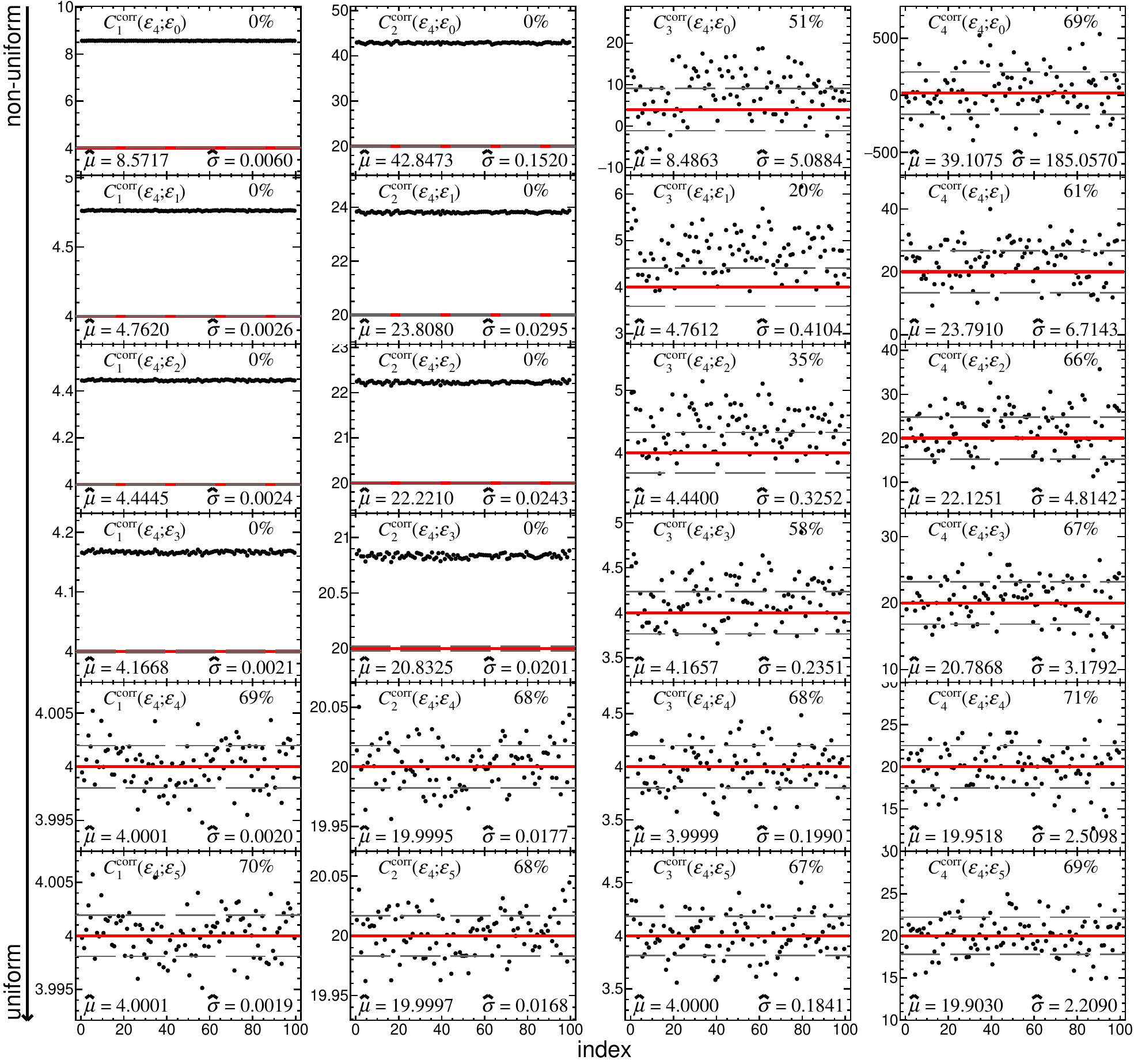}
\caption{(Color online) $C_k^\mathrm{corr}$ ($k$ = 1, 2, 3, 4) measured with $\varepsilon_4$ and corrected with each set of efficiency defined in Eqs.~\eqref{eq12}-\eqref{eq17}. Each panel shows 100 points instead of 1000. $\hat{\mu}$ and $\hat{\sigma}$ at the bottom of each panel show the mean value and the standard deviation of 1000 points. The red solid and gray dashed lines denote $C_k^\mathrm{true}$ and $C_k^\mathrm{true}\pm\hat{\sigma}$, respectively. The number at the top right of each panel represents the fraction of the points falling in between the gray dashed lines.}
\label{fig16}
\end{figure*}

\begin{figure*}[htbp]
\centering
\includegraphics[width=0.8\textwidth]{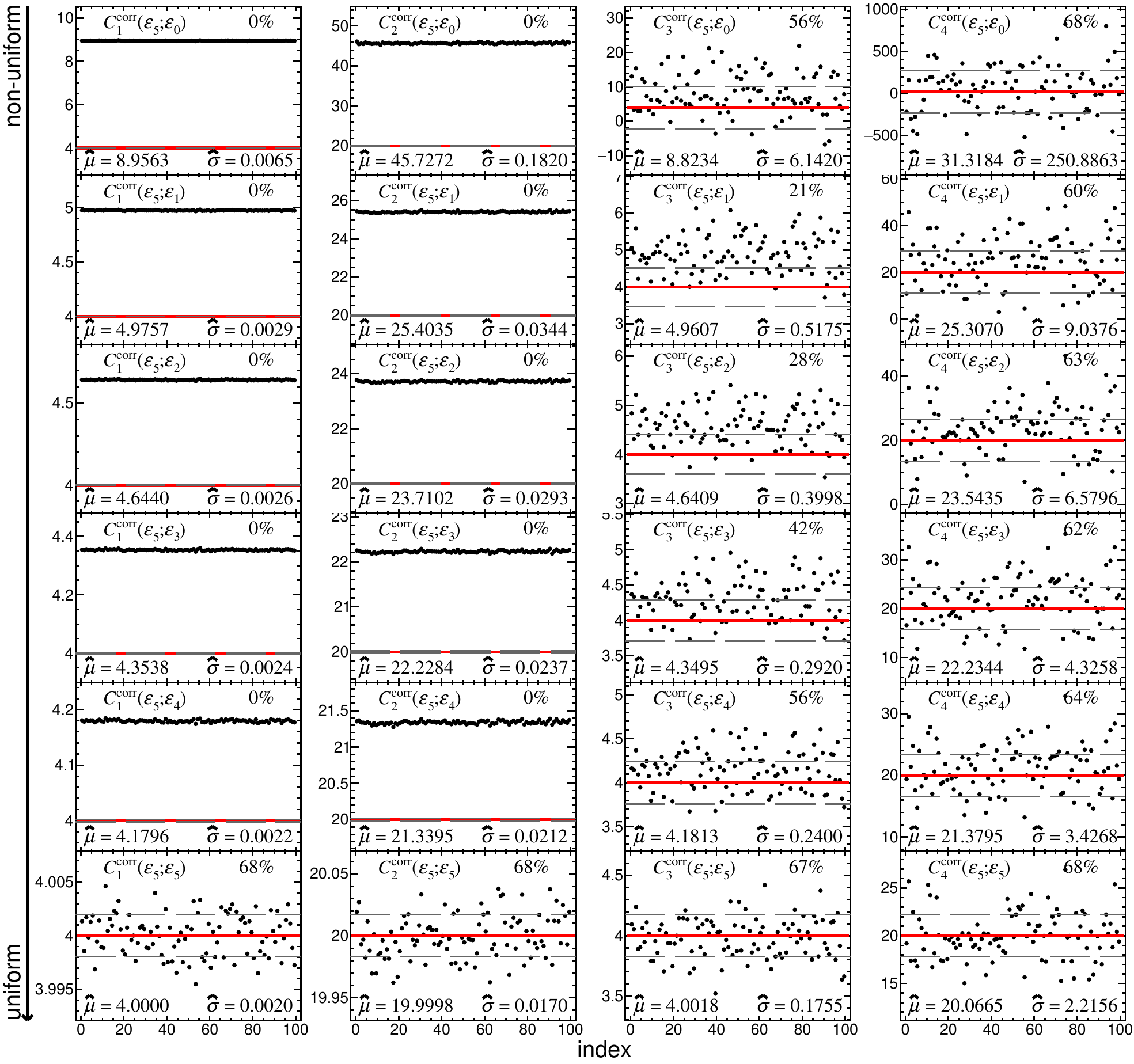}
\caption{(Color online) $C_k^\mathrm{corr}$ ($k$ = 1, 2, 3, 4) measured with $\varepsilon_5$ and corrected with each set of efficiency defined in Eqs.~\eqref{eq12}-\eqref{eq17}. Each panel shows 100 points instead of 1000. $\hat{\mu}$ and $\hat{\sigma}$ at the bottom of each panel show the mean value and the standard deviation of 1000 points. The red solid and gray dashed lines denote $C_k^\mathrm{true}$ and $C_k^\mathrm{true}\pm\hat{\sigma}$, respectively. The number at the top right of each panel represents the fraction of the points falling in between the gray dashed lines.}
\label{fig17}
\end{figure*}

The $C_k^\mathrm{corr}$ ($k$ = 1, 2, 3, 4) measured with $\varepsilon_i$ and corrected with $\varepsilon_j$ ($i$, $j$ = 0, 1, $\cdots$, 5) defined in Eqs.~\eqref{eq12}-\eqref{eq17} are shown in Figs.~\ref{fig12}-\ref{fig17}. $\hat{\mu}$ and $\hat{\sigma}$ shown at the bottom of each panel represent the mean value and the standard deviation of 1000 points, respectively. The red solid lines show $C_k^\mathrm{true}$, and the gray dashed lines denote $C_k^\mathrm{true}$ shifted up and down with $\hat{\sigma}$ ($C_k^\mathrm{true}\pm\hat{\sigma}$). The fraction of $C_k^\mathrm{corr}$ falling in between the gray dashed lines is shown at the top right of each panel.

\end{document}